\documentclass[usenatbib]{mnras}

\usepackage{natbib}
\usepackage{longtable,lscape}
\usepackage{amsmath}
\bibpunct{(}{)}{;}{a}{}{,}
\usepackage{graphicx}

\usepackage{graphicx}
\usepackage{color}
\usepackage{xcolor}
\usepackage{xspace}
\usepackage{ulem}
\newcommand{\MJup}{M$_{\mathrm{Jup}}$\xspace}

\newcommand{\RSun}{R$_{\odot}$\xspace}
\newcommand{\MSun}{M$_{\odot}$\xspace}

\newcommand{\mic}{$\mu$m\xspace}
\newcommand{\as}{\hbox{$^{\prime\prime}$}\xspace}
\title[]{Limits on the presence of planets in systems with debris disks: HD\,92945 and HD\,107146}%\thanks{}}
\author[D. Mesa et al.]{D. Mesa$^{1}$\thanks{E-mail:
    dino.mesa@inaf.it (AVR)}, S. Marino$^{2,3}$, M. Bonavita$^{4,5,1}$,
  C. Lazzoni$^{1,6}$, C. Fontanive$^{7,4,1}$, S. P\'erez$^{8}$, \newauthor
  V. D'Orazi$^{1,9}$, S. Desidera$^{1}$, R. Gratton$^{1}$, N. Engler$^{10}$,
  T. Henning$^{2}$, M. Janson$^{11,2}$, \newauthor
  Q. Kral$^{12}$, M. Langlois$^{13,14}$, S. Messina$^{15}$,
  J. Milli$^{16}$, N. Pawellek$^{3,17}$, C. Perrot$^{18,19,12}$, \newauthor
  E. Rigliaco$^{1}$, E. Rickman$^{20,21}$, V. Squicciarini$^{1,6}$, A. Vigan$^{14}$,
  Z. Wahhaj$^{22,14}$, \newauthor A. Zurlo$^{23,24,14}$,
  A. Boccaletti$^{12}$, M. Bonnefoy$^{16}$, G. Chauvin$^{25,16}$,
  V. De Caprio$^{26}$, \newauthor M. Feldt$^{2}$, L. Gluck$^{16}$,
  J. Hagelberg$^{20}$, M. Keppler$^{2}$, A.-M. Lagrange$^{16}$, R. Launhardt$^{2}$,
  \newauthor A.-L. Maire$^{27}$,  M. Meyer$^{28,10}$, O. Moeller-Nilsson$^{2}$, 
  A. Pavlov$^{2}$, M. Samland$^{2,11}$, \newauthor T. Schmidt$^{12}$, 
  L. Weber$^{20}$ \\ \\
  {\it Affiliations are listed at the end of the paper}
}
\begin{document}
\date{Accepted . Received ; in original form }
\pagerange{\pageref{firstpage}--\pageref{lastpage}} \pubyear{}
\maketitle
\label{firstpage}
\begin{abstract}
  Recent observations of resolved cold debris disks at tens of au have
  revealed that
  gaps could be a common feature in these Kuiper belt analogues. Such gaps
  could be evidence for the presence of planets within the gaps or closer-in
  near the edges of the disk. %Direct imaging can help in putting limits on
  %the mass of these companions.
  We present SPHERE observations of HD\,92945 and
  HD\,107146, two systems with detected gaps. We constrained the mass
  of possible companions responsible for the gap to 1-2~\MJup for
  planets located inside the gap and to less than 5~\MJup for separations down
  to 20~au from the host star. These limits allow us to exclude some of the
  possible configurations of the planetary systems proposed to explain the shape
  of the disks around these two stars. In order to put tighter limits on the
  mass at very short separations from the star, where direct imaging data are
  less effective, we also combined our data with astrometric measurements from
  Hipparcos and Gaia and radial velocity measurements. We were able to limit
  the separation and the mass of the companion potentially responsible for the
  proper motion anomaly of HD\,107146 to values of 2-7~au and 2-5~\MJup,
  respectively.
\end{abstract}


\section{Introduction}
\label{s:intro}

Protoplanetary disks around young (few Myr) stars are nowadays considered
the formation environments for planetary systems
\citep{2012ApJ...756..133C,2014A&A...565A..15M}. In recent years, a growing
number of structures, like e.g. rings, gaps and asymmetries, have been
identified into these disks \citep[e.g. ][]{2015ApJ...808L...3A,
  2017ApJ...837..132V,2018A&A...610A..24F,2018ApJ...859...32P,
  2018ApJ...869L..49I}. While different
mechanisms have been proposed to explain these structures, one of the most
promising is the one implying the presence of planetary mass objects still
in their formation process.
However, at the moment, the detection of planets associated to these structures
has been confirmed for just a handful of cases. The most striking of these
detections is certainly the PDS\,70 system, in which two planetary companions
have been detected through direct imaging in the gap of the disk
\citep{2018A&A...617A..44K,2018A&A...617L...2M,2019NatAs...3..749H,
  2019A&A...632A..25M}. \par
At older ages (from tens to hundreds Myr), debris disks are expected to
have lost a large part of their gaseous component, while their dust component
is continuously replenished by collisions of small bodies orbiting within Kuiper
belt analogues. The presence of planets can shape these disks as demonstrated
by well known cases like $\beta$\,Pic \citep{2009A&A...493L..21L},
HR\,8799 \citep{2008Sci...322.1348M,2010Natur.468.1080M,2016MNRAS.460L..10B,
  2018MNRAS.475.4953R} and HD\,95086 \citep{2013ApJ...779L..26R,
  2015ApJ...799..146S,2016ApJ...822L..29R}.
This is why asymmetries detected in the debris disk are often
  considered hints for the presence of planetary companions
  \citep[see e.g. ][]{2005Natur.435.1067K,2014Sci...343.1490D,
    2015Natur.526..230B}. Similarly, the existence of wide cavities interior
  to cold debris belts or in between warm asteroid belt analogues and cold
  outer belts has been used to argue the presence of multiple planets clearing
  those wide regions of debris, in analogy to the Solar System architecture
  \citep{2008ApJ...672.1196A,2015ApJ...800....5M,2016MNRAS.462L.116S,
    2018A&A...611A..43L,2018MNRAS.480.2757M,2020A&A...639A..54L}.
  Perhaps the most convincing and constraining evidence for planets at
    tens of au is the presence of gaps in broad cold debris disks. There is a
    growing number of \textit{gapped} debris disks, e.g HD\,107146
    \citep{2015ApJ...798..124R,2018MNRAS.479.5423M}, HD\,92945
    \citep{2011AJ....142...30G,2019MNRAS.484.1257M}, HD\,15115
    \citep{2019ApJ...877L..32M,2019A&A...622A.192E} and HD\,206893
    \citep[][ Nederlander et al., submitted]{2020MNRAS.498.1319M} seen with
    ALMA tracing the distribution of mm-sized grains. Scattered light images
    have also revealed the presence of gaps in a few systems, e.g.
    HD\,131835 \citep{2017A&A...601A...7F}, HD\,141569
    \citep{2016A&A...590L...7P}, HD\,120326 \citep{2017A&A...597L...7B} and
    NZ\,Lup \citep{2019A&A...625A..21B}, but these observations trace small
    \mic-sized dust which not always is a good tracer of the true mass
    distribution in the disk (e.g. Wyatt 2006).\par
 One of the simplest interpretations for those gaps is that they are
    caused by single planets orbiting in those gaps. It is well known that a
    planet embedded in a debris disk will carve a gap around its orbit, where
    particles become unstable and are scattered away by the planet. This is
    the so-called chaotic zone \citep{1980AJ.....85.1122W} and its size has
    been studied extensively both exterior to a planet's orbit
    \citep{2006MNRAS.373.1245Q,2015ApJ...798...83N} and interior to it
    \citep{2015ApJ...799...41M}. Roughly speaking, we expect that the gap size
    will have a radial width proportional to $\mu^{2/7}$, where $\mu$ is the
    mass ratio between the planet and the star. Therefore the width of a given
    gap can be directly linked to the mass of the putative planet carving it.
    \par
However, gaps can also be carved by planets interior to the disk inner
  edges. As shown by \citet{2015MNRAS.453.3329P} the secular interaction and
  scattering between a planet on an eccentric orbit and a disk with a similar
  mass further out, can open an asymmetric and wide gap in the disk. Another
  possibility is that two or multiple planets closer-in could place a secular
  resonance within the disk span creating a gap in the disk as well
  \citep{2018MNRAS.479.5423M}. Moreover, as shown more recently by
  \citet{2020arXiv201015617S}, if we consider the non-negligible mass of debris
  disks and its self-gravity, the secular interaction between the disk and
  single eccentric planet closer-in can also place a secular resonance within
  the disk opening an asymmetric gap \citep[see also ][]{2017ApJ...849...98Z}.
  Although HD\,107146's gap is very symmetric and thus unlikely to be carved
  by the alternative scenarios described above \citep{2018MNRAS.479.5423M},
  these could potentially explain the gaps around HD\,92945, HD\,15115 and
  HD\,206893. In fact HD\,206893 is known to host a brown dwarf orbiting at
  11~au, and \citet{2020arXiv201015617S} showed that this could explain the
  gap location given the estimated disk mass for that system.
Another possible scenario to explain gaps in the distribution of
  \mic-sized grains involves planet migration and resonant trapping as proposed
  by \citet{2006ApJ...639.1153W}. This scenario was used by
\citet{2011AJ....142...30G} to explain the structure of the HD\,92945 debris
ring. However, this scenario only works for \mic-sized grains seen in
scattered light and the gap seen also at mm-wavelengths rules it out as
possible mechanism. \par
Tight upper limits on the mass of possible companions, both
inside and outside the gap in the disk, could then allow to constrain these
possible scenarios. This would help understanding the evolutive processes
behind the formation of these structures. While present high-contrast imager
like e.g. SPHERE \citep{2019A&A...631A.155B} or GPI \citep{2006SPIE.6272E..0LM}
are only able to detect planets more massive than 1~\MJup, their data can still
place meaningful upper limits on the masses of the companion carving the gaps.
In this work we present SPHERE observations of two stars hosting debris disks
with gaps, HD\,92945 and HD\,107146, and we use them to put mass limits
for companions within few tens of au from the stars. \par
Furthermore, to complement at shorter separations the limits obtained through
direct imaging, we exploited measurements of the proper motion anomaly (PMa)
obtained by \citet{2019A&A...623A..72K} and radial velocity (RV) data for
these stars. \par
The paper presents a description of the two targets in Section~\ref{s:target}
and describes observations and data reduction in Section~\ref{s:data}. In
Section~\ref{s:result} we describe the results of our analysis while in
Section~\ref{s:discussion} we discuss the results and in
Section~\ref{s:conclusion} we give our conclusion.

\section[]{Target properties}
\label{s:target}
\subsection{HD92945}
HD\,92945 is a K1V \citep{2006A&A...460..695T} star at a distance from the
Sun of 21.54$\pm$0.02~pc \citep{2016A&A...595A...1G,2018A&A...616A...1G}.
%The more recent estimates
%of its age are between 100 and 300~Myrs \citep[e.g. ][]{2004ApJ...614L.125S,
%  2009ApJ...698.1068P,2017A&A...603A...3V}.

\subsubsection{Spectroscopic analysis}
\label{s:spec}

In order to constrain the stellar properties, we analyzed a FEROS
\citep{kaufer99} spectrum from the ESO archive taken on 2018-04-21 under open
time program 0101.A-9012(A) (P.I. R. Launhardt). 
The spectrum covers a wavelength range between $\approx$ 3600 and 9000 \AA
with a nominal resolution of R=48000; the median signal-to-noise ratio (SNR)
per pixel is 260. We carried out continuum normalization and rest-frame correction using the 2020 version of $iSpec$ (\citealt{blanco-cuaresma2014};
\citeyear{blanco-cuaresma2019}); equivalent width (EW) have been measured with ARESv2 \citep{sousa2015}. 
Atmospheric parameters and metallicity ([Fe/H]) have been obtained through
EW measurements of iron lines with the $q^2$ python wrapper
of MOOG (\citealt{sneden73}, 2017 version), developed and maintened by I.
Ramirez \citep{ramirez2014}. We used the ODFNEW ATLAS9 set of model atmopsheres by
\cite{castelli03} and the line list published in \cite{dorazi2020}. 
Our analysis results in the following parameters: $T_{\rm eff}$=5147$\pm$40~K, $\log{g}=4.46\pm0.13$~dex, 
microturbulent velocity $1.15\pm0.12$~km/s and [Fe/H]=$-0.01\pm$0.03~dex.
Broad band colors  (B-V, V-I, Bp-Rp, and V-K) are consistent with the spectroscopic estimate
of about $\approx$5200~K from the tables in \citet{pecaut2013}.

\subsubsection{Stellar age}

\begin{figure*}
\centering
\includegraphics[width=0.5\textwidth,angle=90]{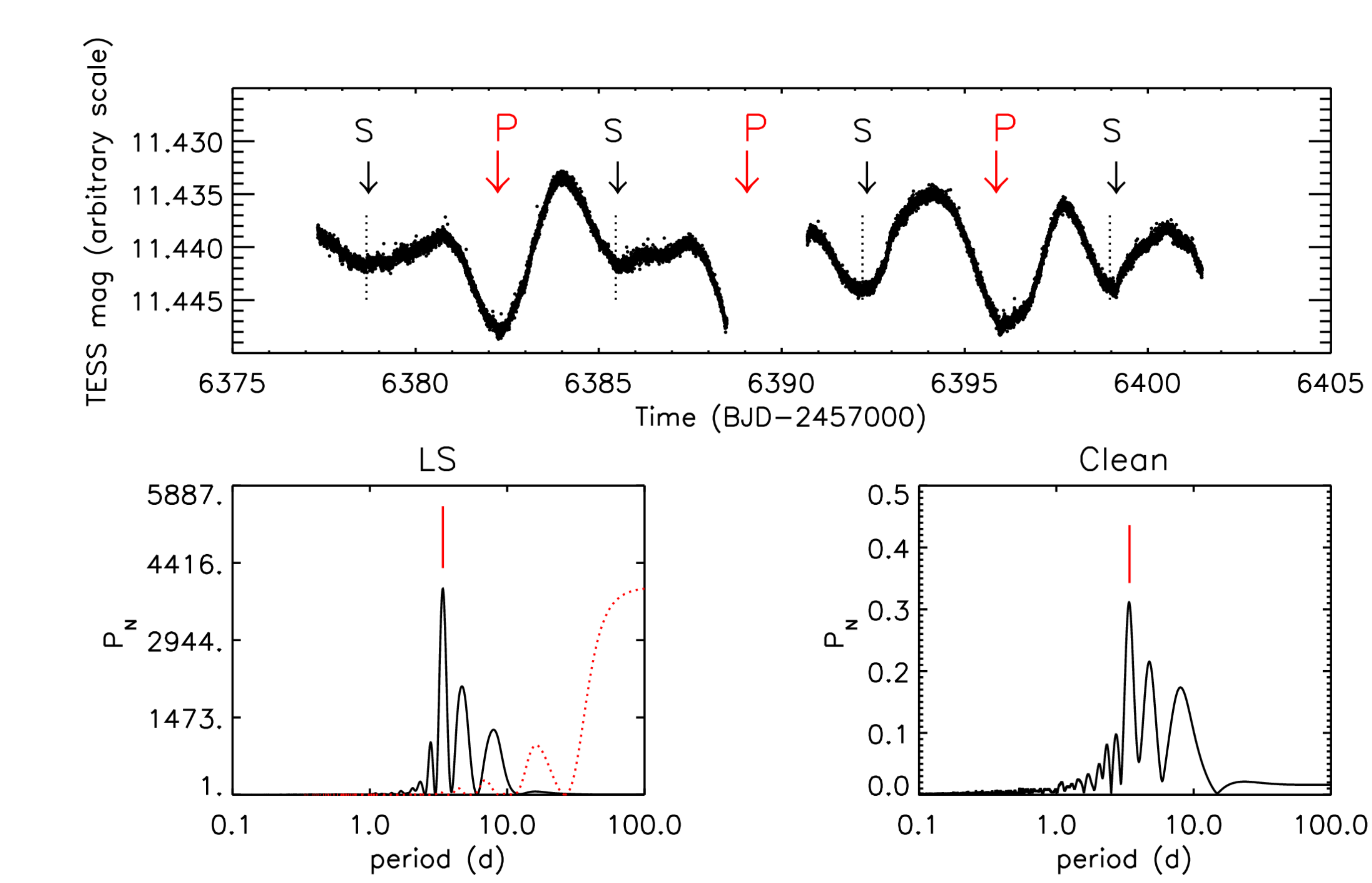}
\caption{Results of periodogram analysis of HD\,92945 with TESS data. 
{\it Top panel: }  magnitudes vs. barycentric Julian Day. 
  Red and black downward arrows mark the position of primary (P) and secondary (S) spots, respectively, in a double dip light curve. Vertical dotted lines mark the epochs of observed secondary minima, which are receding with respect to the computed epochs of minima (black arrows). {\it Bottom panels:} Lomb-Scargle (left) and Clean (right) normalized periodograms (P$_{\rm N}$). The red dotted line shows the spectral window
  function superimposed on the LS periodogram and the solid vertical lines mark the power peaks at half rotation period.}
\label{f:HD92945_periodogram}
\end{figure*}

There are various recent estimates of the age of the star, typically, 
between 100 and 300~Myr
\citep[e.g.][]{2004ApJ...614L.125S,2009ApJ...698.1068P,2017A&A...603A...3V}.
\citet{nielsen2019}  quote membership to the AB Dor with an estimated age
  of $149^{+31}_{-49}$~Myr \citep{2017A&A...603A...3V,2015MNRAS.454..593B}.
We reconsidered the available age indicators. \par 
We searched for the stellar rotation period by exploiting the photometric variability  in the TESS data
\citep{tess}. We considered the TESS pre-search data conditioning simple aperture photometry (PDCSAP) time series, removing data in the time interval 1555--1557 (BJD-2457000), where residual instrumental effects were significant. 
The period search was carried out computing the Lomb-Scargle periodogram \citep{Press02,Scargle82} and the Clean \citep{Roberts87} periodogram, which has the capability of effectively
removing possible beat frequencies arising from the data sampling.
As shown in the bottom panels of Figure~\ref{f:HD92945_periodogram}, the Lomb-Scargle (LS)
periodogram  revealed a major power peak at 
$P=3.40\pm0.24$~days (while the Clean procedure shows P = 3.39$\pm$0.24~days), with a False Alarm Probability (FAP) of the order of 10$^{-6}$. 
However, a
visual inspection of the top panel of Figure~\ref{f:HD92945_periodogram} shows the lightcurve to be double dip, owing to spots on opposite
hemispheres. Therefore, the rotation period is likely P = 6.8$\pm$0.34~days. This value is in agreement with the P = 7.176\,d derived using the
period-activity relation from \citet{noyes1984}.  We marked with red and black downward arrows the epochs of the primary (P) and secondary (S) minima computed according to a rotation period P = 6.8\,d.
We note a significant evolution of the secondary spot, which produces progressively deeper minima from one rotation cycle to the next, as well as it shows a slight migration towards earlier epochs, as shown by the position of the vertical dotted lines with respect to black downward arrows. \rm
A photometric period which is roughly two times our value was published by
\citet{strassmeier2000}.
We also note that our rotation period is fully compatible with the other age
indicators and with the observed projected rotational velocity, at odds with a
1/2 $\times$ or 2$\times$ period. 
The FAP associated with our detected period, which is the probability that a
peak of given height in the periodogram is caused simply by statistical variations, i.e., Gaussian noise, was computed through
Monte Carlo simulations, i.e., by generating 1000 artificial light
curves obtained from the real light curve, keeping the date but
permuting the magnitude values \citep[see, e.g., ][]{Herbst02}.
We followed the
method used by \citet{Lamm04} to compute the errors associated with the period determination. \rm
\par
We found that the rotation period falls close but slightly above the sequence
of the Pleiades \citep{rebull2016}, and between the rotation period
distributions of similar color stars of M35 and M34 open clusters 
\citep[ages 150 and 220 Myr, respectively, ][]{meibom2009,meibom2011}. \par
The chromospheric emission ($\log{R^{\prime}_{HK}}=-4.32$; \citealt{wright2004}) is within
the distribution of the Pleiades, while the X-ray emission
($\log{L_{X}/L_{bol}}=-4.47$) is below the Pleiades locus. The corresponding
ages using \citet{mamajek2008} calibrations are 160 and 310 Myr, respectively.
\par
The equivalent width of the Li 6708\AA\,doublet measured on the FEROS spectrum
is 153.2$\pm$1.2~m\AA, within the observed distribution of Pleiades members but
sligthly below the mean values for the cluster. \par
Kinematic analysis using the BANYAN\footnote{\url{http://www.exoplanetes.umontreal.ca/banyan/banyansigma.php}} $\Sigma$ online tool \citep{gagne2018}
yield null membership probability for AB Dor and other young moving groups.
\par
All these finding are compatible with an object with age close but sligthly
older than the Pleiades. From a weighted combination of the available indicators \citep{desidera2015}, we adopt 170 Myr with age limits 120 to 250 Myr.

\subsubsection{Mass, radius, and inclination}
\label{s:paramHD92945}
The stellar mass derived using the PARAM\footnote{\url{http://stev.oapd.inaf.it/cgi-bin/param_1.3}} interface \citep{param2006}, adopting the
spectroscopic effective temperature and metallicity, and considering only the
age range for the indirect methods as in \citet{desidera2015}, is
0.86$\pm$0.01~\MSun.
A stellar radius of $0.75\pm0.02$~\RSun is also derived in the fit. 
Coupling it with the observed rotation period and the projected rotational
velocity $v\sin{i}=4.5\pm1.0$~km/s \citep{nordstrom2004,valenti2005}, we infer
an inclination of $53.7^{+31.3}_{-12.4}$~deg. \rm
This is compatible with the inclination of the disk (see below), although the
uncertainties also allow a significant misalignment.

\subsubsection{The debris disk}

The presence of a debris disk around this star was first inferred by
\citet{2000PhDT........17S} based on the excess at 60~\mic measured by
IRAS. Mid-infrared and submillimiter data obtained using Spitzer and
  the Caltech Submillimiter Observatory revealed excess also at 70~\mic and
  350~\mic consistent with a a disk optically thin with an inner radius of
  15-23~au and a temperature of 40-46~K \citep{2005ApJ...634.1372C,2009ApJ...698.1068P}.
The disk was resolved for the first time by
\citet{2011AJ....142...30G} using {\it Hubble space telescope} (HST) images in
the V and I bands. They also estimated the fractional infrared luminosity of
the disk to be $7.7\times10^{-4}$. Recently, \citet{2019MNRAS.484.1257M}
observed its disk using ALMA at 0.86~mm and detected, into the disk of
planetesimals that extends from 50 to 140~au, a wide ($20_{-8}^{+10}$~au) gap at
a separation of 73$\pm$3~au from the star. They also found for the disk an
inclination of $65.4\pm0.9^{\circ}$ and a position angle of $100.0\pm0.9^{\circ}$.
They concluded that if the gap was caused by a planet on a circular orbit, the
planet would need to be less massive than 0.6~\MJup. Based on
a possible asymmetry in the disk, they also proposed that the gap could have
been carved due to secular resonance caused by two planets interior to the disk.
In this context, they also defined different possible masses for these two
objects. \par
Because this system has a bright debris disk, it has been target of a large
number of direct imaging surveys \citep[e.g. ][]
{2013ApJ...773...73J,2013ApJ...777..160B,2017A&A...603A...3V}.  In particular,
\citet{2013ApJ...777..160B}, adopting an age of 100~Myr and exploiting NICI
near-infrared observations and the AMES-COND \citep{2003IAUS..211..325A}
evolutionary models, were able to put a mass limit of $\sim$10~\MJup at a
separation of 0.5\as ($\sim$10.8~au), of $\sim$4~\MJup at a separation of
1\as ($\sim$21.5~au) and less than 2~\MJup at separation larger than 2\as
($\sim$43~au).

\subsection{HD\,107146}
HD\,107146 is a G2V \citep{1970AJ.....75..507H} star at a distance from the
Sun of 27.47$\pm$0.03~pc \citep{2016A&A...595A...1G,2018A&A...616A...1G}.
%Its estimated age is between 80 and 200~Myr \citep{2004ApJ...604..414W}.
\subsubsection{Spectroscopic analysis}

We exploited one FEROS spectrum (with a median SNR per pixel of 183) available
from the ESO archive and obtained on 2018-04-30 under open time program
0101.A-9012(A) (P.I. R. Launhardt). 
Following the same approach described in Section \ref{s:spec}, we obtained
$T_{\rm eff}=5933\pm53$ ~K, $\log{g}= 4.56\pm0.09$~dex, microturbulent velocity
$1.19\pm0.08$~km/s and [Fe/H]=+0.01$\pm$0.03~dex. \par
Broad band colors (B-V, V-I, Bp-Rp, and V-K) are consistent with the spectroscopic estimate of about
5900~K obtained from \citet{pecaut2013} tables.

\subsubsection{Stellar age}

\begin{figure*}
\centering
\includegraphics[width=0.5\textwidth,angle=90]{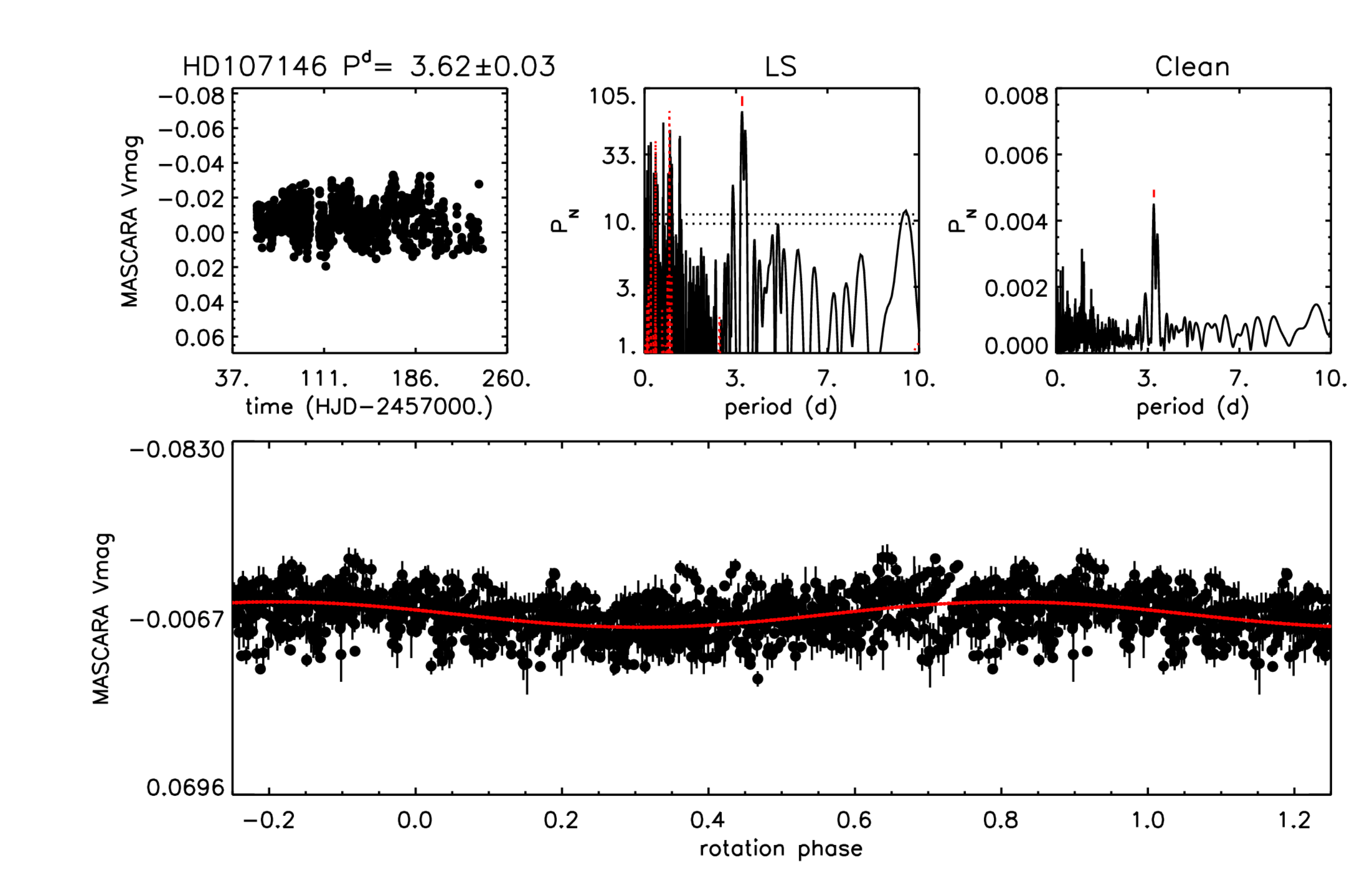}
\caption{Results of periodogram analysis of HD\,107146 using MASCARA data.
    In the top left panel we plot magnitudes vs heliocentric Julian Day. In the
    top middle panel we plot the Lomb-Scargle periodogram with the spectral
    window function and power level corresponding to FAP=1\% overplotted (red
    dotted line), and we indicate the peak corresponding to the rotation
    period. In the top right panel we plot the CLEAN periodogram. In the bottom
    panel we plot the light curve phased with the rotation period. The solid
    line represents the sinusoidal fit.}
\label{f:HD107146_periodogram}
\end{figure*}

HD\,107146 was observed as target of the MASCARA \citep[The Multi-site All-Sky
  CAmeRA;][]{Burggraaff2018} project. Observations were collected from
February 2015 until April 2016, in two consecutive seasons for a total of
7188 magnitude measurements, with a cadence of about 1 measurement every 10
minutes.  After removal of outliers at 3$\sigma$ level, we binned the data
with a bin width of 1h and we were left for the subsequent analysis with 1716
average magnitudes with an average precision of $\sigma$=0.0045\,mag.
Both Lomb-Scargle and Clean periodogram analyses revealed the same rotation
period within the uncertainties in the complete time series as well as in the
single observation seasons. In Figure~\ref{f:HD107146_periodogram}, we
summarize the results for the first observation season whose monitoring turned
out to be more homogeneous with no intra-season gaps.
We found a rotation period of P=3.62$\pm$0.03~days with a
False Alarm Probability (FAP) $<$1\% and an amplitude of the magnitude
rotational modulation $\Delta$V=0.01\,mag. \par
A comparison with the rotation period distribution of Pleiades members from
\citet{rebull2016} shows the HD107146 period to fit very well into a 125~Myr
period distribution (assuming V=7.04~mag and K=5.54~mag V-K=1.5~mag
and P=3.62~days). Adopting the rotation-age relations by \citet{mamajek2008} we
derive for HD\,107146 an age of 156~Myr. \par
The activity indicators yield consistent results: the chromospheric emission 
\citep[$\log R_{HK}=-4.34$; ][]{wright2004} is within the distribution of the
Pleiades, while the X-ray emission ($\log L_{X}/L_{bol}=-4.33$) is slightly below
the Pleiades locus. The corresponding ages using the \citet{mamajek2008}
calibrations are 189 and 226 Myr, respectively. \par
We measured on the FEROS spectrum a  Li equivalent width  of
$125.3\pm2.0$~m\AA\, basically identical to the determination by
\citet{wichmann2003} (125 m\AA). This collocates HD\,107146 within the lower
envelope of the Pleiades distribution. \par
Moreover, the search for an association between HD\,107146 and young
  moving groups did not give any result. We also searched on Gaia DR2
  archive to identify possible comoving objects within $1^{\circ}$ from the star
  without identifying any.
%Finally, the star does not result member of known young moving groups and a
%dedicated search on Gaia DR2 does not identify comoving objects within 1 deg.
\par 
On the basis of rotation period, level of chromospheric and coronal activity,
and Lithium content, the age of HD107146 appears to be comparable to or
slightly older than the Pleiades age. Also in this case we calculated a
  weighted combination of the available indicators following
  \citet{desidera2015}, and adopted an age of 150~Myr with
limits between 120 to 200~Myr.

\subsubsection{Mass, radius, and inclination}

The stellar mass derived using the same method used for HD\,92945 as described
in Section~\ref{s:paramHD92945} is 1.06$\pm$0.01~\MSun. A stellar radius of
$0.94\pm0.02$~\RSun is also derived in the fit. Coupling it with the observed
rotation period and the projected rotational velocity $v\sin{i}=5.0\pm0.5$~km/s
\citep{nordstrom2004,valenti2005}, we infer an inclination of
$22.4^{\circ}\pm5.5^{\circ}$. This inclination is well aligned with the inclination
estimated for the disk (see below) within the error bars.

%%% COMMENTO POTREBBE ESSERE IL CASO DI FARE UNA TABELLA CON TUTTI I PARAMETRI STELLARI (+eventualmente quelli del disco)

\subsubsection{The debris disk}

The debris disk was first discovered by \citet{2000PhDT........17S} using IRAS
observations at 60~\mic while observations at sub-millimeter wavelenghts
confirmed the presence of dust \citep{2004ApJ...604..414W}. Later, the disk was
resolved thanks to HST observations \citep{2004AAS...20512708A,
  2011A&A...533A.132E} revealing a disk extending up to a separation of
$\sim$160~au and with a brightness peak at 120~au. The infrared fractional
luminosity of the disk was estimated to be $1.07\times10^{-3}$
\citep{2011A&A...533A.132E}.
Spitzer observations at 24~\mic and 70~\mic allowed to determine the
  presence of a warm dust belt in the system with a temperature of
  $\sim$120~K and located at a separation of 5-15~au from the star
\citep{2011ApJ...730L..29M,2014MNRAS.444.3164K}.
Using ALMA data at 1.25~mm,
\citet{2015ApJ...798..124R} found that the disk extends from 30 to 150~au
from the star but with a decrease of brightness at intermediate radii that,
in their interpretation, could be due to the presence of an earth mass planet
orbiting at 80~au from the star. Finally, \citet{2018MNRAS.479.5423M} observed
HD\,107146 with ALMA at 0.86 and 1.1~mm and found that the planetesimal belt
extends from 40 to 140~au and identified a circular gap at separations
from 60 to 100~au with a drop in emission by $\sim$50\%.
%Moreover, they
%detected an inner extended emission peaked at $\sim$20~au that they attributed
%to the presence to a inner warm belt.
Finally, they were able to constrain the disk inclination to $19.3\pm1.0^{\circ}$
and the position angle to $153.0\pm3.0^{\circ}$ and concluded that either
multiple low mass planets or a single planet migrating through the disk were
needed to make a wide gap that is only 50\% empty.  

%\par
  
\section[]{Observations and data reduction}
\label{s:data}

\subsection{Direct imaging data}
\label{s:dataimaging}

\begin{table*}
 \centering
 \begin{minipage}{150mm}
   \caption{Characteristics of the SPHERE observations presented in this work.
     In column 1 we list the observing night, in column 2 we list the target
     name. In column 3 and 4 we report the SPHERE observing mode and the
     coronagraph used, respectively. In columns
     5 and 6 we list the number of datacubes, the number of frames for each
     datacube and the exposure time for each frame for observation with IFS and
     with IRDIS, respectively. In column 7 we report the total rotation of the
     FOV during each observation. In columns 8, 9 and 10 we report
     the median values of seeing, coherence time and wind speed during the
     observations. \label{t:obs}}
  \begin{tabular}{c c c c c c c c c c}
  \hline
  Date & Target & OBs. Mode & Coronagraph &  Obs. IFS  & Obs. IRDIS & FOV rot. & seeing & $\tau_{0}$  & wind \\
       &        &       &       &      &    &($^{\circ}$) & (\as) & (ms)  & (m/s)  \\
    \hline
         2018-01-27  & HD\,92945  & IRDIFS & N\_ALC\_YJH\_S  &  16$\times$6;48  &  16$\times$6;48  & 121.3 &  0.80  &  5.6  &  6.63 \\
         2015-04-09  & HD\,107146 & IRDIFS &N\_ALC\_YJH\_S  &  16$\times$5;32  &  16$\times$10;16 & 14.9  &  1.66  &  3.8  &  0.80 \\
         2016-03-19  & HD\,107146 & IRDIFS & N\_ALC\_YJH\_S  &  16$\times$5;32  &  16$\times$10;16 & 15.7  &  1.00  &  2.1  &  4.35 \\
         %2018-04-22  & HD\,107146 & IRDIS DPI & N\_ALC\_YJH\_S  &     not used     & 26$\times$4;32  & 21.6  &  0.47  &  7.2 &   2.25 \\
\hline
\end{tabular}
\end{minipage}
\end{table*}

\begin{figure*}
\centering
\includegraphics[width=\textwidth]{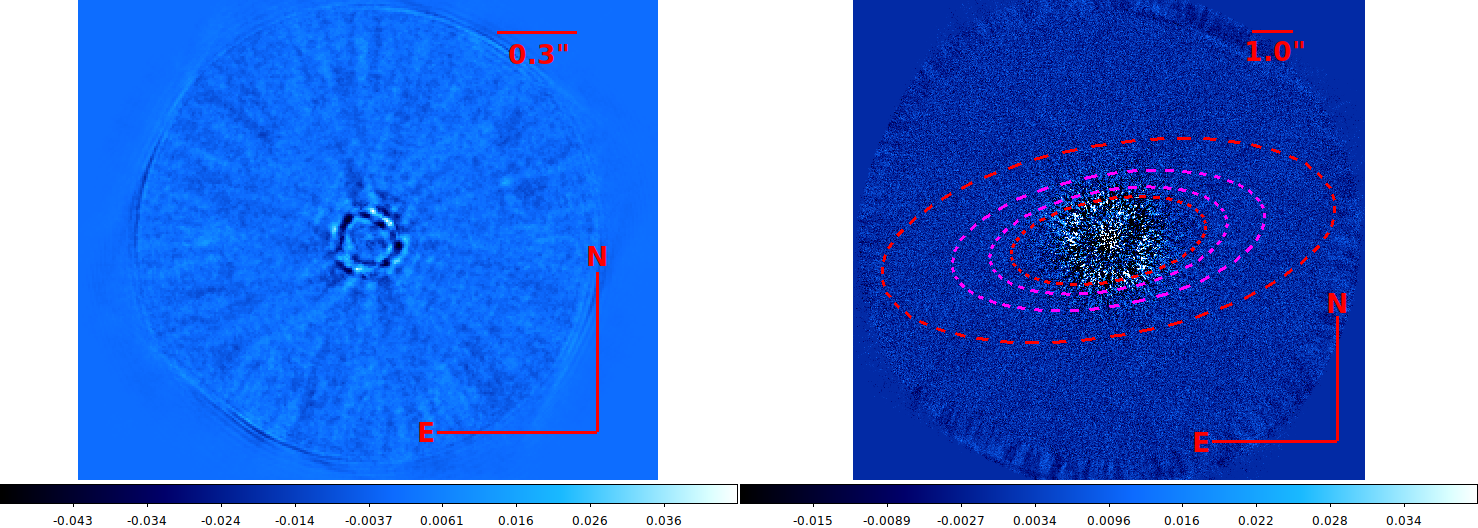}
\caption{Final images obtained for HD\,92945 using IFS data (left panel) and
  IRDIS (right panel). In both cases the data were reduced using PCA 
    subtracting 100 and 25 principal components for IFS and IRDIS,
    respectively. On the
  IRDIS image we also plotted dashed lines indicating the position of the outer
  belt of the disk (red lines) and of the gap (magenta lines). All these lines
  are outside the FOV of IFS.}
\label{f:HD92945img}
\end{figure*}

\begin{figure*}
\centering
\includegraphics[width=\textwidth]{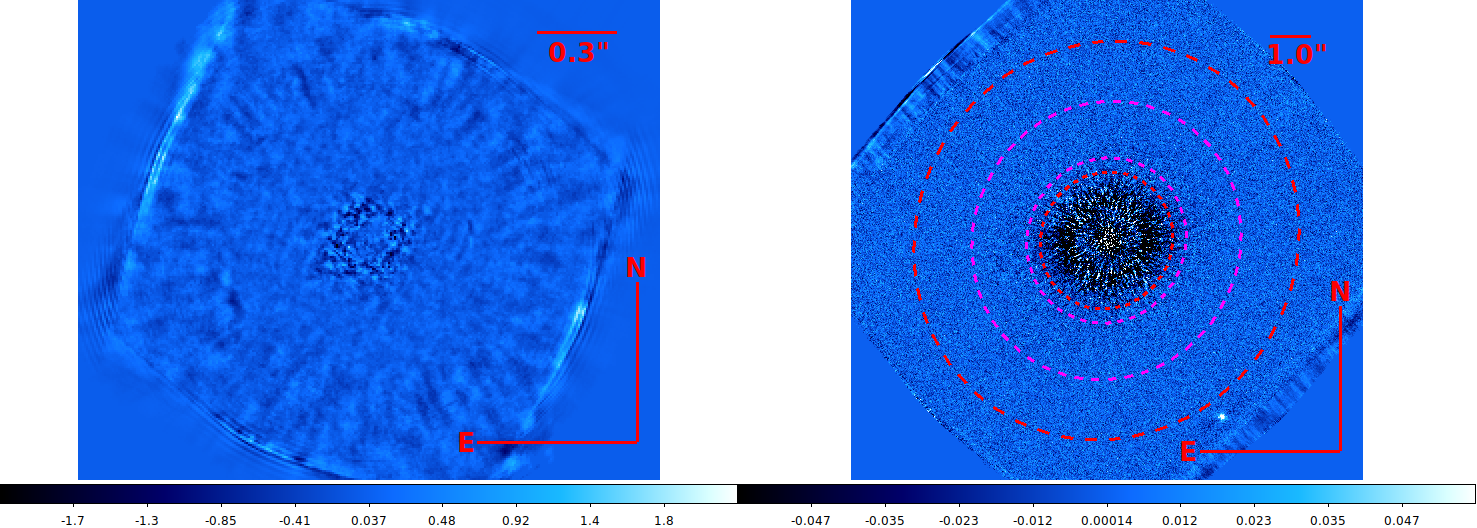}
\caption{Final images obtained for HD\,107146 using IFS data (left panel) and
  IRDIS (right panel) for the 2016-03-19 epoch. In both cases the data were
  reduced using PCA subtracting 100 and 25 principal components for IFS
    and IRDIS, respectively. We also plotted dashed lines to indicate the
  position of the outer belt of the disk (red lines) and of the gap (magenta
  lines) as defined in \citet{2018MNRAS.479.5423M}. These lines are outside
  the IFS FOV.}
\label{f:HD107146img}
\end{figure*}

HD\,92945 was observed during the 2018-01-27 night in the context of the
SHINE survey \citep{2017sf2a.conf..331C,2020arXiv200706573V} with SPHERE
operating in IRDIFS observing mode, that is with IFS \citep{2008SPIE.7014E..3EC}
observing in Y and J spectral bands between 0.95 and 1.35~\mic while IRDIS
\citep{2008SPIE.7014E..3LD} observed in the H spectral band with the H23 filter
pair \citep[wavelenght H2=1.593~\mic; wavelength H3=1.667~\mic; ][]{2010MNRAS.407...71V}. \par
HD\,107146 was observed in two different epochs in the context of the open
time program 095.C-0374(A) (P.I. L.~Ricci). The two epochs were acquired on
2015-04-09 and 2016-03-19, respectively. In both cases the observations were
carried on with SPHERE operating in the IRDIFS observing mode. The observations
for both targets were performed in pupil stabilized mode to be able to implement
angular differential imaging \citep[ADI; ][]{2006ApJ...641..556M}. The main
characteristics of all the observing nights are listed in Table~\ref{t:obs}.
\par
For all the observations, frames with satellite spots symmetric with respect to
the central star were obtained before and after the coronagraphic sequences.
This allowed to determine the position of the star behind the coronagraphic
focal plane mask and accurately recenter the data. Furthermore, to allow
to correctly calibrate the flux of companions, images with the star not behind
the coronagraph were taken. In these cases, the use of an appropriate neutral
density filter was mandatory to avoid saturation of the detector. \par
The data were reduced through the SPHERE data center
\citep{2017sf2a.conf..347D} that perform the requested calibrations, both for
IFS and IRDIS, using the data reduction and handling
\citep[DRH; ][]{2008ASPC..394..581P} pipeline. On the reduced data we then
applied speckle subtraction algorithms using both TLOCI
\citep{2014SPIE.9148E..0UM} and principal components analysis
\citep[PCA; ][]{2012ApJ...755L..28S} as implemented in the
SPHERE consortium pipeline called SpeCal
\citep[Spectral Calibration; ][]{2018A&A...615A..92G}. \par
%Beside to these data we also retrieved for HD\,107146 IRDIS dual-polarization
%imaging \citep[PDI; ][]{2014SPIE.9147E..1RL,2020A&A...633A..63D,2020A&A...633A..64V} mode data obtained in the context of the open time program 0101.C-0843(A)
%(P.I. S. Perez). The observations were performed during the night of 2018-04-22
%exploiting the H broad spectral band. These data were reduced using the
%IRDIS data reduction for accurate polarimetry
%\citep[IRDAP; ][]{2017SPIE10400E..15V,2020A&A...633A..64V} pipeline appositely
%created to reduce the SPHERE/IRDIS polarized data.

\subsection{Radial velocity data}
\label{s:rvdata}

To complement the direct imaging mass limits at lower separations from the
star we retrieved archival RV data for both stars.
For HD\,92945 we retrieved from the ESO archive data obtained using the
HARPS spectrograph \citep{2006SPIE.6269E..0PL} operating at the 3.6~m
telescope at La~Silla. These data were obtained in the context of the open time
programs 074.C-0037(A)(P.I. E. Guenther), 075.C-0202(A) (P.I. E. Guenther) and
192.C-0224(B) (P.I. A.-M. Lagrange). The data were reduced using the official
HARPS pipeline (DRS v. 3.5), that delivers the RV and full width at half
maximum (FWHM) measured on each Cross-Correlation Function (CCF). The mask used
is G2. We gathered a total of 28 RVs, ten of them being taken in 2005, the
remaining ones in 2016-17. Part of the data was then taken after an instrument
upgrade in June~2015 \citep{2015Msngr.162....9L} and display an offset
toward larger RVs with respect to the pre-upgrade data. This offset was
applied before combining data. The typical precision for these observations
was of the order of 1~m/s. \par
RVs data for HD\,107146 covering the epochs from 2007 to 2015 were obtained
using the SOPHIE high-resolution echelle spectrograph fiber-fed from the
Cassegrain focus of the 1.93 m telescope at the Haute-Provence Observatory
(OHP) in France \citep{2006tafp.conf..319B}. SOPHIE is installed in a
temperature-stabilized environment, and the dispersive elements are kept at
constant pressure to provide high-precision radial velocities
\citep{2008SPIE.7014E..0JP}. Since June 2011, a new fiber scrambler has
provided a significant improvement of the spectrograph illumination stability,
leading to a precision gain of a factor $\sim$6 \citep{2011SPIE.8151E..15P,
  2013A&A...549A..49B}. The spectra were reduced and extracted using the
SOPHIE pipeline \citep{2009A&A...505..853B}, and the resulting wavelength
calibrated 2D spectra were correlated using a numerical binary mask
corresponding to spectral type G2 to obtain the radial velocity measurement
\citep{1996A&AS..119..373B}. In this case the typical precision was of the
order of 2~m/s.

\subsection{Proper motion data}
\label{s:datadeltamu}

The presence of a companion can be inferred studying the anomalies
of the proper motion of the host star. These anomalies are obtained
  comparing the long-term proper motion vector derived comparing data from the
  Hipparcos \citep{2007A&A...474..653V} and the Gaia DR2 catalogues and the
  instantaneous proper motion values both at the Hipparcos and Gaia DR2 epochs.
  Deviations between these vectors at S/N levels higher than 3 are indication
  of the presence of a companion.
Even in the case of no evidence of a
low mass companion, these data can however be used to define constraints on the
mass of a companion especially at low separations from the host star where
the limits from direct imaging are less effective or totally absent due to the
presence of the coronagraph. In a recent work, \citet{2019A&A...623A..72K}
calculated the PMa for a sample of 6741 stars at a
distance less than 50~pc from the Sun. Both HD\,92945 and HD\,107146 were part
of this sample and we can then exploit these results in combination with the
limits from direct imaging and from RV to constrain the presence of companions
around these systems. While for the case of HD\,92945 they did not find any
evidence of a companion, for HD\,107146 they found a signal to noise (S/N)
of 3.42 on the PMa strongly indicative of the presence of a companion. We will
discuss in Section~\ref{s:discussion} the implications of this result.

\section{Results}
\label{s:result}

\subsection{HD\,92945}
\label{s:HD92945res}

\begin{figure}
\centering
\includegraphics[width=\columnwidth]{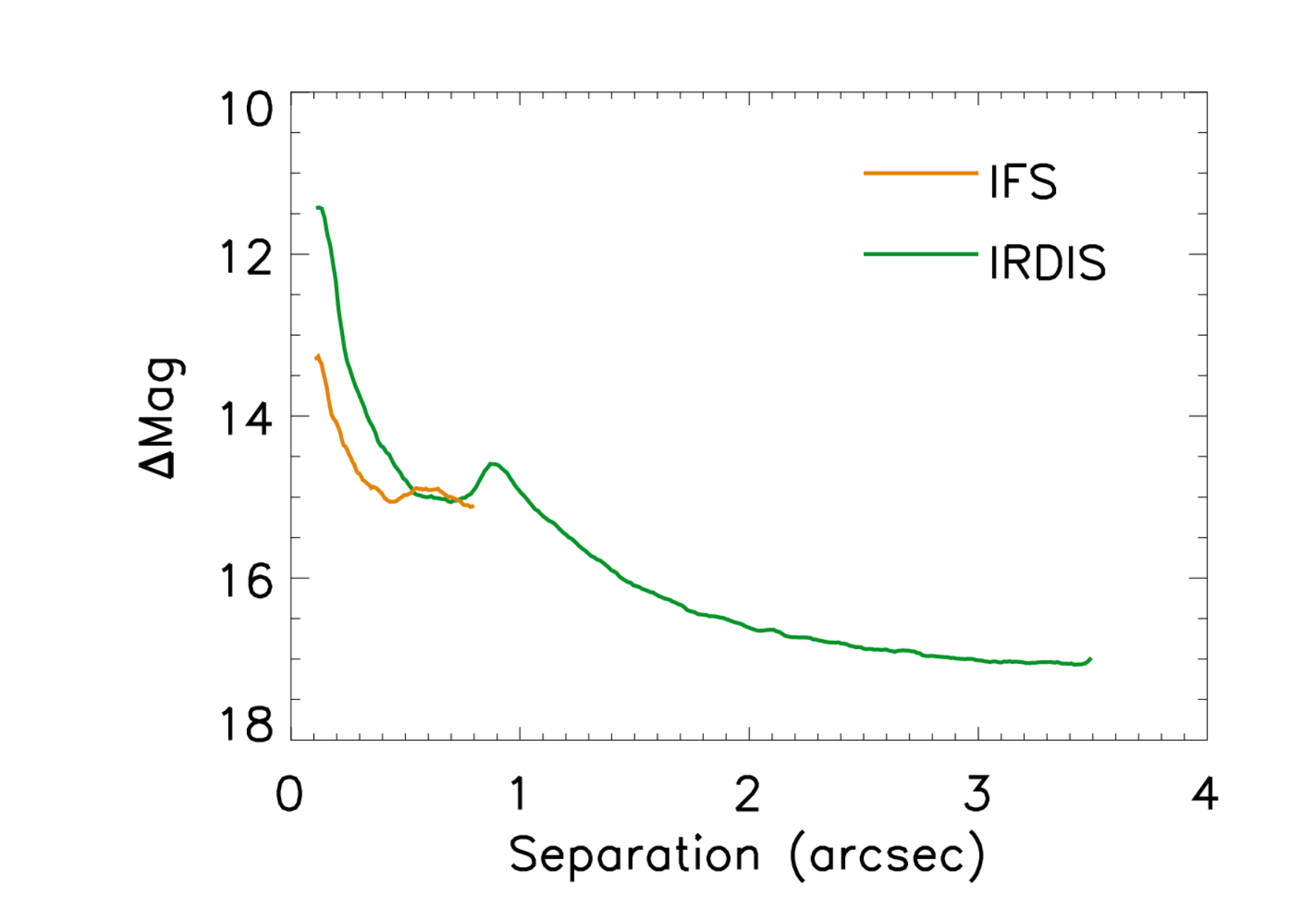}
\caption{Magnitude contrast versus projected separation in the region around HD\,92945 both for IFS (orange line) and IRDIS (green line).}
\label{f:HD92945contrast}
\end{figure}

\begin{figure}
\centering
\includegraphics[width=\columnwidth]{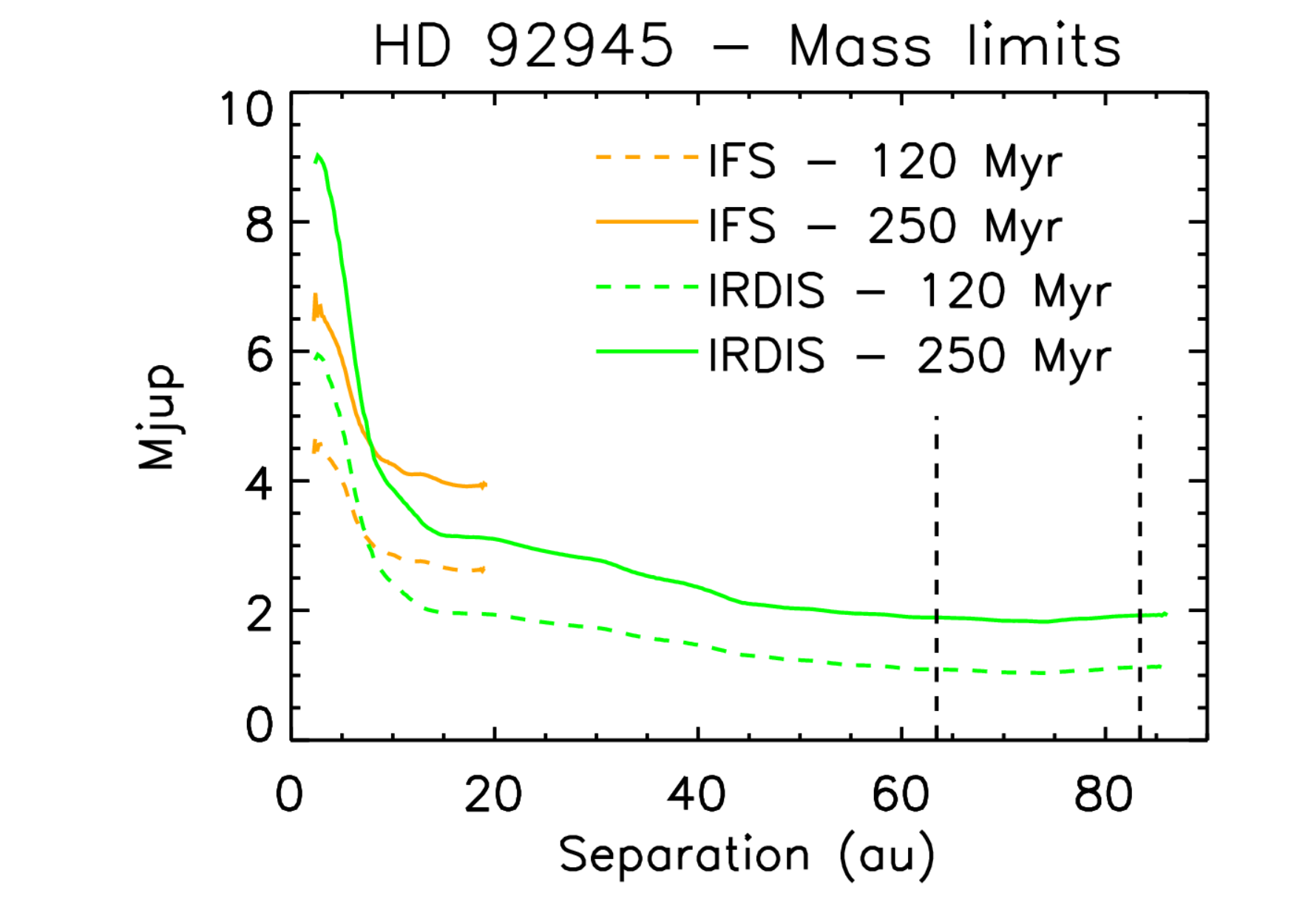}
\caption{Mass limits obtained for HD\,92945 using the AMES-COND models both
  for IFS (orange lines) and for IRDIS (green lines). The dashed lines
  represents the mass limits obtained assuming for the system an age of
  120~Myr, the solid lines are obtained assuming an age of 250~Myr. The two
  dashed vertical lines correspond to the expected position for the gap.}
\label{f:HD92945mass}
\end{figure}

%\begin{figure}
%\centering
%\includegraphics[width=\columnwidth]{HD92945_masslimplot.png}
%\caption{Mass limits obtained converting the IFS (orange) and IRDIS (blue)
%  $\Delta$mag limits for HD\,92945 to mass, using the AMES-COND models. The
%  width of the curves shows the difference in mass limits obtained assuming an
%  age of 100~Myr (lower edge) and 300~Myrs (upper edge). The gray area shows
%  the position of the belt.}
%\label{f:HD92945mass}
%\end{figure}

%\begin{figure}
%\centering
%\includegraphics[width=\columnwidth]{HD92945_DMU+DI.png}
%\caption{Output of COPAINS for HD\,92945 using Gaia DR2 (blue) and Hipparcos
%  (purple) as short-term proper motion measurements, and Tycho-2 for long-term
%  proper motions. The solid lines and shaded areas correspond to the median and
%  1$\sigma$ intervals, assuming a flat distribution in eccentricity. The black
%  dashed and solid lines represent the limits obtained from direct imaging
%  observation assuming for the system an age of 100 and 300~Myr, respectively.}
%\label{f:HD92945deltamu}
%\end{figure}

In Figure~\ref{f:HD92945img} we display the final images obtained for 
HD\,92945 using both IFS (left panel) and IRDIS (right panel). We
overplotted on the IRDIS image dashed red line to indicate the position
of the disk and dashed magenta lines to indicate the position of the gap as
indicated by \citet{2019MNRAS.484.1257M}. The same positions were however
outside the field of view (FOV) of IFS. No clear detection of possible
candidate companions is in the FOV of both instruments. Also, the disk is
undetected as expected due its low surface brightness typical for $\sim$100~Myr
old debris disk and ADI self-cancellation effects \citep{2012A&A...545A.111M}.
%and the moderately low inclination of the disk with respect to line of sight.
We then calculated the $5\sigma$ luminosity contrast as a function of the
separation from the star adopting the technique presented in
\citet{2015A&A...576A.121M} and corrected for the small sample statistic
following the method described by \citet{2014ApJ...792...97M}. 
The contrast in magnitude derived with this method both for IFS and IRDIS is
displayed in Figure~\ref{f:HD92945contrast}. With IFS, we obtain a contrast of
the order of 15 magnitudes at separations larger than 0.5\as, while IRDIS
allows to obtain contrasts better than 16 magnitudes at separations larger than
$\sim$1.5\as. These contrast estimates assume that the disk is optically thin
which is true for this system and for debris disks in general. \par
Adopting this contrast and exploiting the AMES-COND evolutionary models, we
defined the mass limits
for possible companions around HD\,92945 adopting both the minimum (120~Myr)
and the maximum (250~Myr) age proposed for this system. The separation on
  the disk plane was considered in this case adopting an inclination of
  $65.4^{\circ}$ and a position angle of $100.0^{\circ}$
  \citep{2019MNRAS.484.1257M}. Using these informations we calculated for each
  position in the image both the mass limits and the deprojected separation from
  the star. The median value of the mass for positions at the same separation
  from the star was assumed as the mass limit at that separation. The results
obtained from this procedure are displayed in Figure~\ref{f:HD92945mass} where
the dashed lines represent the mass limits obtained for the 120~Myr age case
while the solid lines represent the 250~Myr case. Also, in this Figure we
distinguish between the IFS case, represented with the two orange lines, and the
IRDIS case, represented with the two green lines. The two dashed vertical
  lines, finally, represent the expected position of the gap in the disk.
From our results, we can see that at this position the mass limits are between
1 and 2~\MJup depending from the age of the system. These limits are still
higher than the maximum mass of 0.6~\MJup evaluated for a single planet carving
the gap proposed by \citet{2019MNRAS.484.1257M}. Nevertheless, these limits are
useful to rule out the presence of massive companions interior or exterior to
the disk. As for the solution with two inner planets proposed by
\citet{2019MNRAS.484.1257M} to explain the gap (see e.g. Figure~8 in their
paper), our data are able in this case to further constrain the possible
configurations proposed in that paper. \par
%The results obtained through the proper motion offsets using the method
%described in Section~\ref{s:datadeltamu} are presented in
%Figure~\ref{f:HD92945deltamu}, for Gaia~DR2 (blue) and Hipparcos (purple), and
%assuming a primary mass of 0.77 \MSun. The solid lines and shaded areas
%correspond to the median and 1$\sigma$ intervals of possible mass and
%separation pairs compatible with the weak proper motion trends observed for
%the two stars. We assumed a Gaussian distribution in eccentricity from
%\citet{2013PASP..125..849B}, and the resulting solutions are marginalised over
%all other orbital parameters. Figure~\ref{f:HD92945deltamu} we also overplot 
%black lines indicating the detection limits from our SPHERE observations,
%adopting the upper and lower age estimates of our systems.
%From this comparison we can see that the limits from the proper motion offset
%are generally higher than those obtained through direct imaging apart that at
%very short separations (less than few au) where the direct imaging cannot
%provide any limit due to the presence of the coronagraph. 

%The lack of detection in the imaging data allows us to rule out the outer parts of the COPAINS dynamical constraints, suggesting that any companion to either star would need to be below $\sim$10 M$_\mathrm{Jup}$ and within 1--2 AU.

\subsection{HD\,107146}
\label{s:HD107146res}

\begin{table*}
 \centering
 \begin{minipage}{100mm}
   \caption{Astrometric positions relative to the host star of the candidate
     companion detected around HD\,107146. \label{t:astroHD1017146}}
  \begin{tabular}{c c c c c}
  \hline
  Obs. date & $\Delta\alpha$ (mas) & $\Delta$Dec (mas) & Separation (mas) & PA (deg) \\
    \hline
         2015-04-09  &  -2948.58$\pm$4.00 &  -4487.17$\pm$4.00 & 5369.25$\pm$5.66 & 213.3$\pm$0.5 \\
         2016-03-19  &  -2811.68$\pm$4.00 &  -4315.67$\pm$4.00 & 5150.78$\pm$5.66 & 213.1$\pm$0.5 \\
\hline
\end{tabular}
\end{minipage}
\end{table*}

\begin{figure}
\centering
\includegraphics[width=1.3\columnwidth]{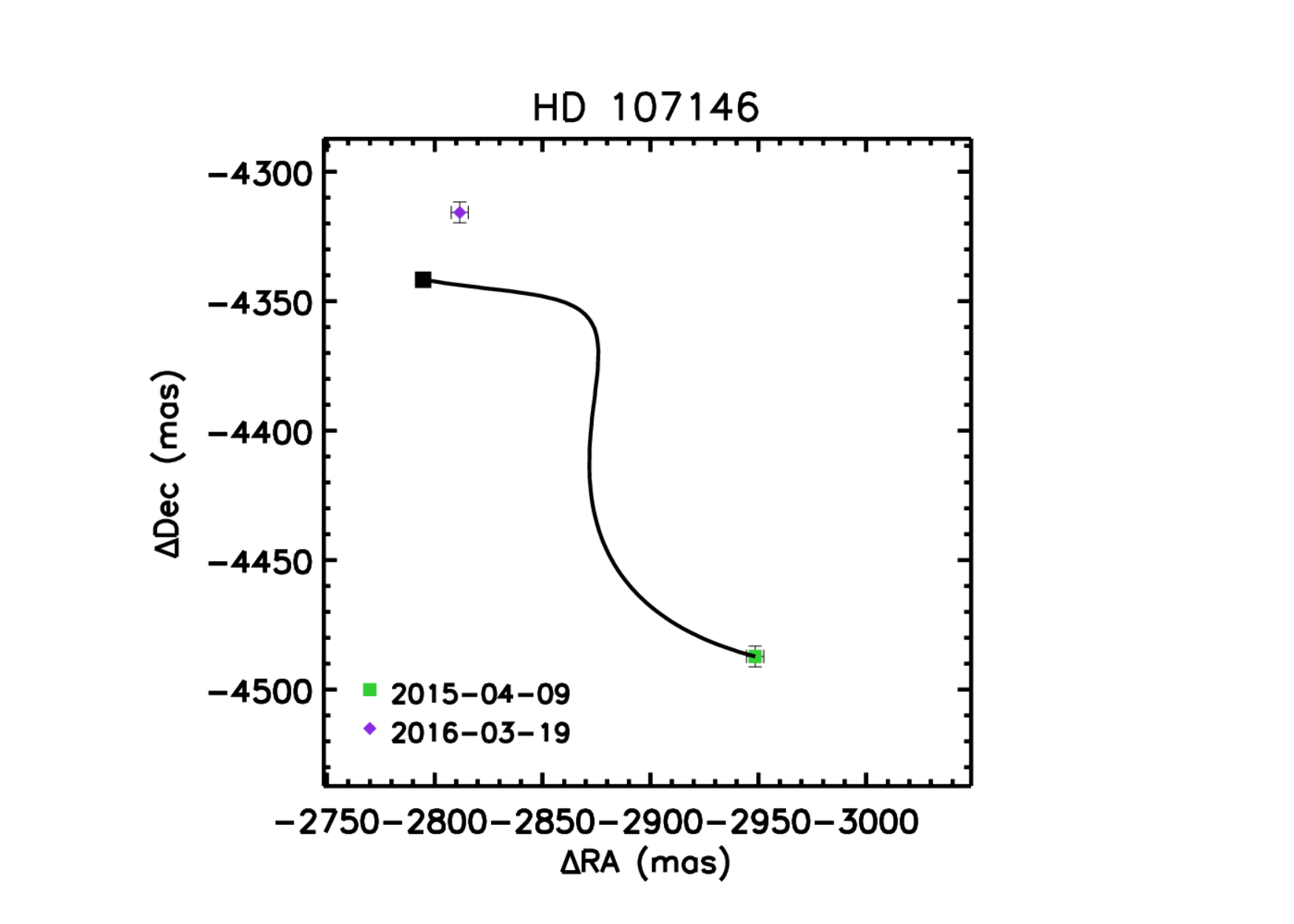}
\caption{Relative astrometric positions of the candidate companion of
  HD\,107146 with respect to the host star in the two observing epochs.
  The green square represents the relative position of the candidate
    companion at the first observing epoch. The violet diamond represents the
    relative position of the candidate companion in the second observing
    epoch. The solid black line represents the
  expected course of the companion if it were a background object. The black
  square at the end of the line represents the expected position at the epoch
  of the second observation in this latter case.}
\label{f:HD107146astro}
\end{figure}

\begin{figure}
\centering
\includegraphics[width=\columnwidth]{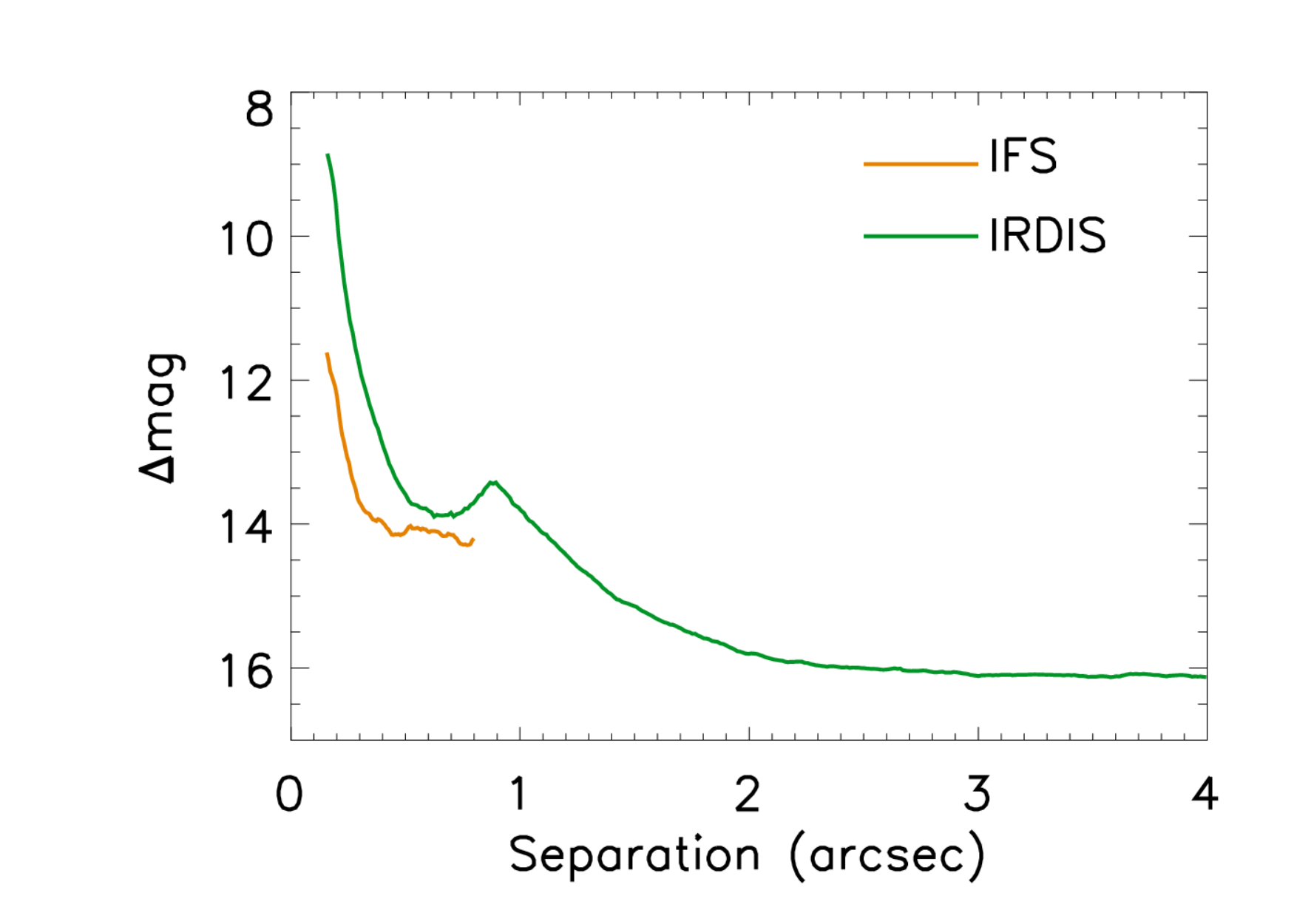}
\caption{Magnitude contrast versus projected separation in the region around HD\,107146 both for IFS (orange line) and IRDIS (green line).} %and IRDIS DPI (blue line)
  %data.}
\label{f:HD107146contrast}
\end{figure}

\begin{figure}
\centering
\includegraphics[width=\columnwidth]{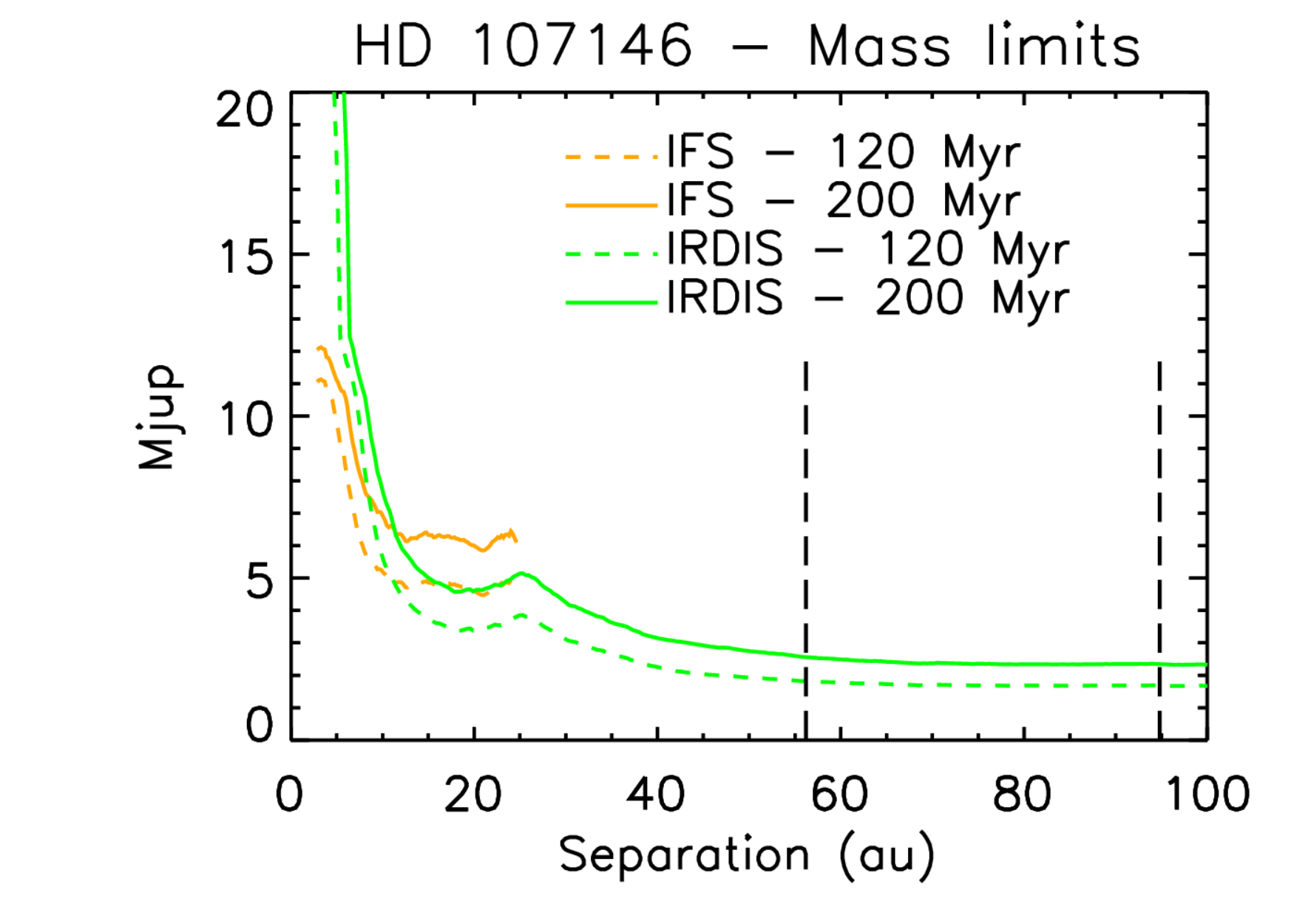}
\caption{Mass limits obtained for HD\,107146 using the AMES-COND models both
  for IFS (orange lines) and for IRDIS (green lines). %and for IRDIS/DPI (blue
  %lines).
  The dashed lines represents the mass limits obtained assuming for
  the system an age of 120~Myr, the solid lines are obtained assuming an age
  of 200~Myr. The two dashed vertical lines correspond to the expected position
  for the gap.}
\label{f:HD107146mass}
\end{figure}

%\begin{figure}
%\centering
%\includegraphics[width=\columnwidth]{HD107146_DMU+DI.png}
%\caption{Output of COPAINS for HD\,107146 using Gaia DR2 (blue) and Hipparcos
%  (purple) as short-term proper motion measurements, and Tycho-2 for long-term
%  proper motions. The solid lines and shaded areas correspond to the median and
%  1$\sigma$ intervals, assuming a flat distribution in eccentricity. The black
%  dashed and solid lines represent the limits obtained from direct imaging
%  observation assuming for the system an age of 80 and 200~Myr, respectively.}
%\label{f:HD107146deltamu}
%\end{figure}

In Figure~\ref{f:HD107146img} we display the final images obtained both for
IFS (left panel) and for (IRDIS) for the HD\,107146 data obtained during
the night of 2016-03-19. We also overplotted on the IRDIS image red dashed
lines to indicate the position of the disk and magenta dashed lines to indicate
the position of the gap using the values obtained by
\citet{2018MNRAS.479.5423M}. These lines are outside the IFS FOV. Like for
the case of HD\,92945 we are not able
to retrieve any signal from the disk. Also in this case, this is expected
due to the low luminosity at estimated age of the system and to the moderately
low inclination of the disk with respect to line of sight. In this case,
however, a point source is clearly detected at a separation larger than 5\as
and position angle of $\sim$213$^{\circ}$ at south west of the star in both the
observing epochs (see Table~\ref{t:astroHD1017146} for precise astrometric
data). %This position is particularly interesting because it would
%be just outside the external edge of the planetesimal belt. 
In Figure~\ref{f:HD107146astro} we display the comparison of the relative
astrometric positions in the two epochs
listed in Table~\ref{t:astroHD1017146}. The relative position of
the candidate
in the second epoch (violet diamond), is very similar to its expected position
if it was a background object (black squares) while we would expect a
  reduced shift from the relative position of the first epoch (green square)
  if it were a bound object. From this analysis we then
conclude that the candidate is very likely a background object. \par
Following the same method described in Section~\ref{s:HD92945res} for
HD\,92945 we calculated the luminosity contrast in the region around
HD\,107146. The results of this procedure are shown in
Figure~\ref{f:HD107146contrast} both for IFS (orange line) and IRDIS (green
line). %and for the IRDIS DPI data%(blue line).
From IFS we obtain a contrast better than 14~magnitudes at
separations larger than 0.5\as while IRDIS allows to obtain a contrast better
than 16~magnitudes at separations larger than 2\as. \par %In any case, we obtain
%the best contrast using the IRDIS\,DPI data with a contrast of the order of
%15~magnitudes at a separation of 0.5\as and around 16.5~magnitudes at
%separations larger than 2.4\as. \par
In Figure~\ref{f:HD107146mass} we display the results of the procedure that,
using the AMES-COND evolutionary models and assuming for the disk an
inclination of $19.3^{\circ}$ and a position angle of $153^{\circ}$
\citep{2018MNRAS.479.5423M}, converts these contrast limits
into mass limits adopting both a minimum age of 120~Myr (dashed lines) and
a maximum age of 200~Myr (solid lines). The two dashed vertical lines
represent the expected position for the gap in the disk. In the gap the mass
limits that we obtain from our data are between 1 and 2.5~\MJup. At lower
separations, we reach limits slightly lower than 5~\MJup in between 20-50~au.
Interior to 20~au the mass limits are typically larger than 5~\MJup. \par
%Like done for the HD\,92945 case, we present the result results obtained for
%HD\,107146 through the proper motion offsets using the method in
%Figure~\ref{f:HD107146deltamu}. For the HD\,107146 case we assumed a primary
%mass of 1.09 \MSun. In the case the Tycho-2 - Gaia DR2, $\Delta\mu$ is
%consistent with 0 at the 1$\sigma$ level, which causes the lower part of the
%parameter space to be filled by the blue shade for this target. We note that
%the results from COPAINS are consistent within 1$\sigma$ interval with the
%limits obtained by \citet{2019A&A...623A..72K} exploiting proper motion data
%from Hipparcos and Gaia-DR2 catalogues. Also in
%Figure~\ref{f:HD107146deltamu} we overplot with black lines the limits obtained
%through direct imaging observations. Also in this case the limits from proper
%motion offsets improve the results from direct imaging at short separations
%from the star (less than 10~au) giving limits of the order of 5-10~\MJup.

\section{Discussion}
\label{s:discussion}
\subsection{Comparison with dynamical models}
\label{s:dynmod}

To further check how the limits we obtained can help to constrain the
structures of these systems we compared them with the results of dynamical
models following the method devised in \citet{2018A&A...611A..43L}. This
method derives analytical tools to link the extension of gaps detected in debris
disks with the masses, semi-major axes and eccentricities of putative planets
responsible for the gaps. The main assumption underneath this formulation is
that a planet sweeps dust particles from a region around its orbit called
chaotic zone. The width of the chaotic zone is directly proportional to the
mass, the semi-major axis and the eccentricity of the planet. When more than
one planet is present in the gap, not only the planet-disk interaction has to
be taken into account but also the planet-planet interaction must be treated.
Since a stable system is needed to preserve the debris disk, we included this
condition in deriving the multi-planetary architecture. Moreover, to avoid
further degeneracies in masses and semi-major axes, we assumed the planets
with equal mass and as close as possible to have still a marginally stable
system. We will refer to this last assumption as max packing condition. \par 
For what regards the circular orbit case, the planet-disk interaction is ruled
by the equation derived in \citet{2015ApJ...799...41M}. See Equations~3 and 4
in \citet{2018A&A...611A..43L} for more details. Moreover, for two planets on
circular orbits the stability criterium was derived from
\citet{1993Icar..106..247G} setting the eccentricities to zero and assuming
equal masses to simplify the calculation as shown by Equation~23 in
\citet{2018A&A...611A..43L}. We also considered the case of three equal mass
planets. In this case the stability of the system is guaranteed by the the
Hill criterium coupled with K-parametrization curves derived by
\citet{2014MNRAS.442.1110M} and are also reported in Equation~ 25 in
\citet{2018A&A...611A..43L}. \par
We decided to not consider the case of eccentric orbit because the disk
eccentricity for both the system is very low with an upper limit of 0.097 for
HD\,92945 \citep{2018MNRAS.479.5423M} and of 0.03 for HD\,107146
\citep{2019MNRAS.484.1257M}. This should exclude the presence of
high-eccentricity embedded planets \citep{PearceWyatt2014,2015MNRAS.453.3329P}. 
%For the case of eccentric orbits, the planet-disk interaction is described 
%either by \citet{1980AJ.....85.1122W} for the case when $e_p<0.3$ or by
%\citet{2009MNRAS.399.1403M} when $e_p\geq0.3$. \par
%In their equations, we adopted as semi-major axis the pericenter and
%apocenter value the planet (eq 9,10,11 and 12 Lazzoni+18).
%In this case, the planet can shape a wider area, getting closer to the outer disk at its apocenter and to the inner one at the pericenter. However, eccentric planets are expected to force a similar eccentricity on the disk (Mustill&Wyatt+2009) so that, in the cases analyzed in this paper, we can confidently exclude highly eccentric solutions.
For all the considered cases we assumed that the gap was completely devoid
from dust particles. Other studies \citep[e.g. ][]{2016MNRAS.462L.116S}
consider the gap as only marginally cleared. In such cases, the masses derived
for the planets will be lower with respect to the ones derived in our analysis. We can then consider our solutions as upper limits for each architecture
considered. \par

Thanks to the procedure described above, we estimated that, in the case of
HD\,92945, a single planet on a circular orbit should have a mass of
at most $0.3$~\MJup to carve the gap. This is slightly lower but however
in good agreement with the upper limit of 0.6~\MJup found by
\citet{2019MNRAS.484.1257M}. If the gap was carved by multiple planets, their
mass could be even lower. Clearly, all these planets have masses well below
the detection limit of around 1~\MJup reachable with SPHERE or with any other
present planet imager. \par
In the case of HD\,107146 the gap could be caused by a single planet with a
mass of at most 3.3~\MJup. This would allow the detection according
to the mass limit we obtained for this system. We have however to stress that,
as explained above, this value has to be regarded just an upper limit. %In any
%case this result reinforces the conclusion by \citet{2018MNRAS.479.5423M} that
%multiple planets are required to explain its formation.
Considering two or three planets would require a mass of 0.47~\MJup
and 0.03~\MJup, respectively, for all the considered planets. This is 
well outside the detection capability of our observations. \par
%We also considered (for the case with one or two planets) the possibility of
%eccentric orbits. However, this always results in masses well below the
%detection limits of our instrument. As an example of this analysis we display
%in Figure~\ref{f:twoplanecc} the dependence of the planetary mass from the
%eccentricities of the orbits of the planets in the case of two planets in the
%gap. 

\begin{figure*}
\centering
\includegraphics[width=\textwidth]{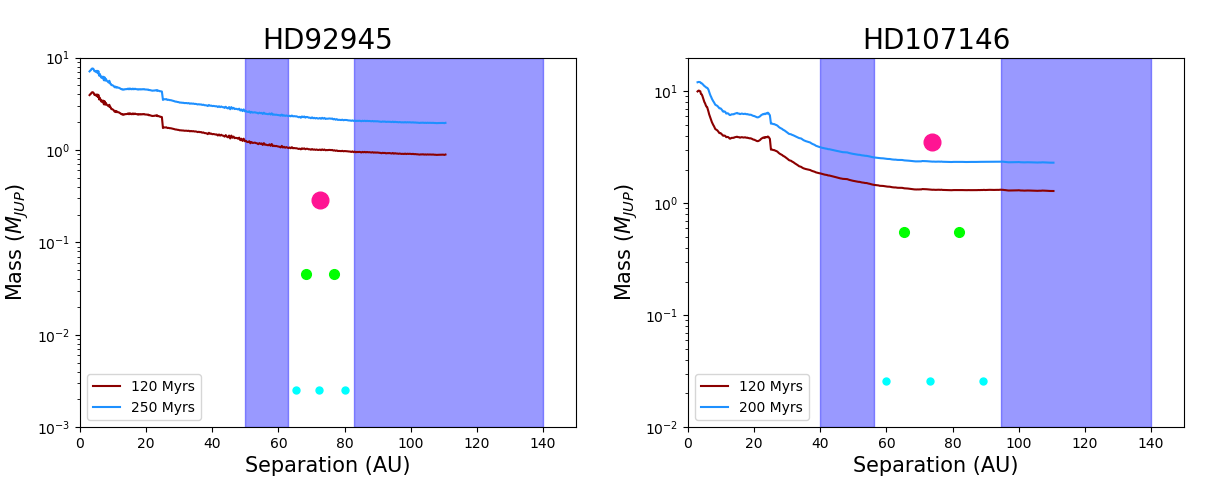}
\caption{Upper limits of planetary masses from dynamical models considering one
  (magenta circle), two
  (green circles) or three (cyan circles) planets with circular orbits both in
  the case of HD\,92945 (left panel) and HD\,107146 (right panel). The shaded
  areas represent the positions of the disk with the gap positions also
  displayed. For comparison we also include the mass limit obtained through
  direct imaging.}
\label{f:oneplan}
\end{figure*}

%\begin{figure*}
%\centering
%\includegraphics[width=\textwidth]{double_planet_eccentric_all.png}
%\caption{Dependence of the planetary mass from the eccentricity of the two
%  planets in the case the gap is carved by two eccentric planets both for
%  HD\,92945 (left panel) and for HD\,107146 (right panel). The color scale on
%  the right of the two plots is expressed in unit of \MJup.}
%\label{f:twoplanecc}
%\end{figure*}

\subsection{Comparison with limits from RV and PMa}
\label{s:complim}
\subsubsection{SPHERE detection probability}

\begin{figure*}
    \centering
    \includegraphics[height=5.75cm]{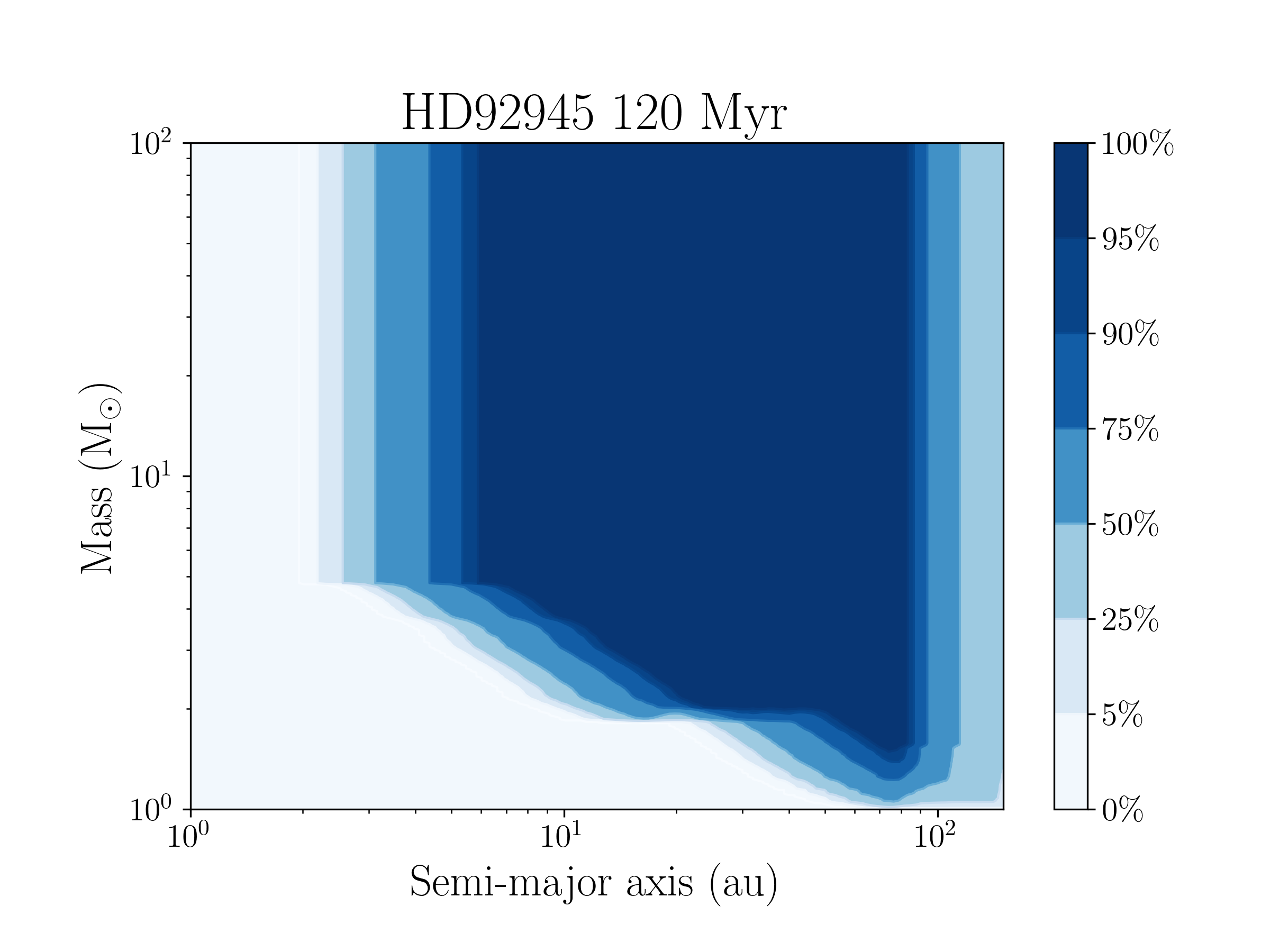}
    \includegraphics[height=5.75cm]{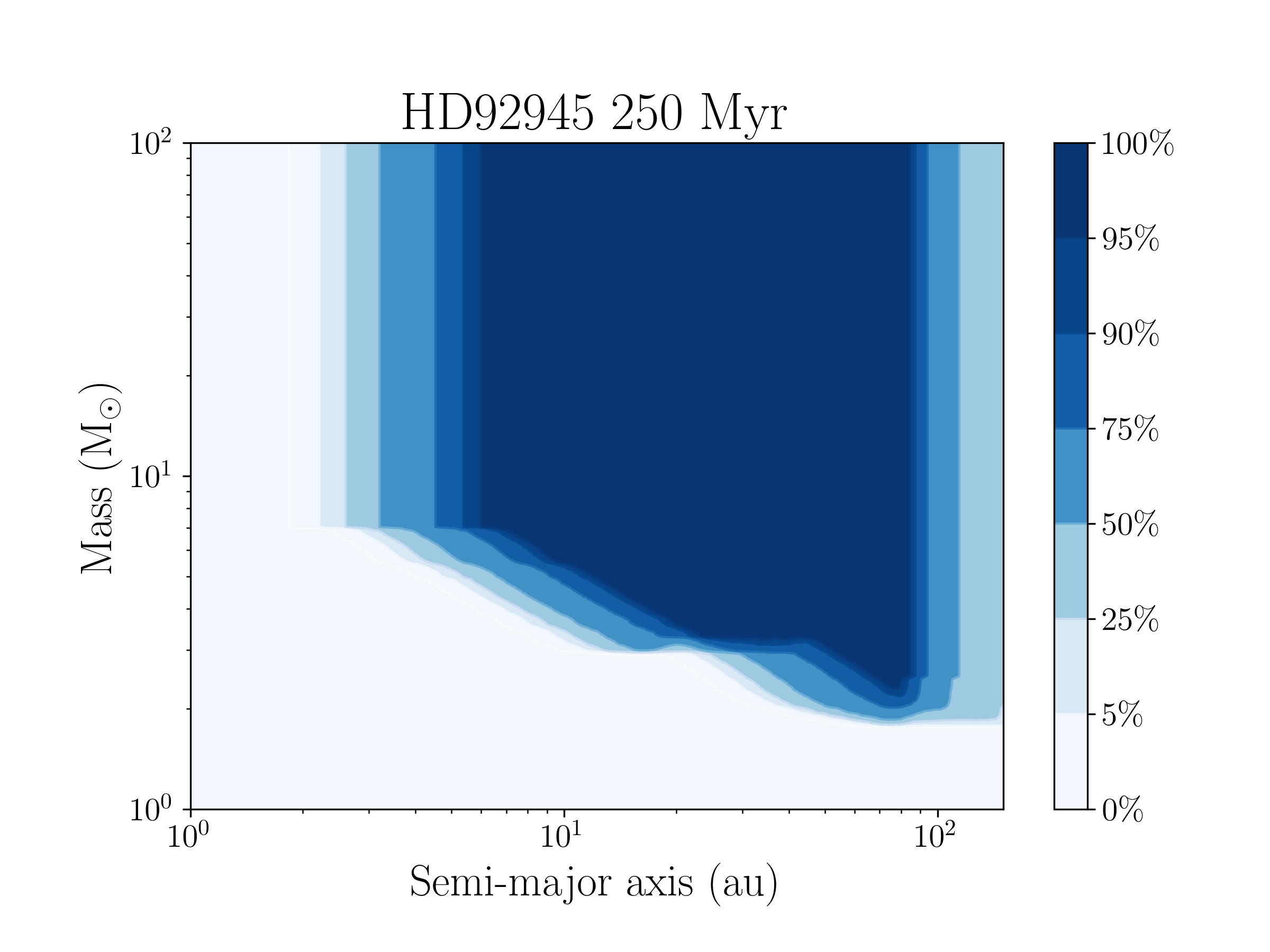}
    \includegraphics[height=5.75cm]{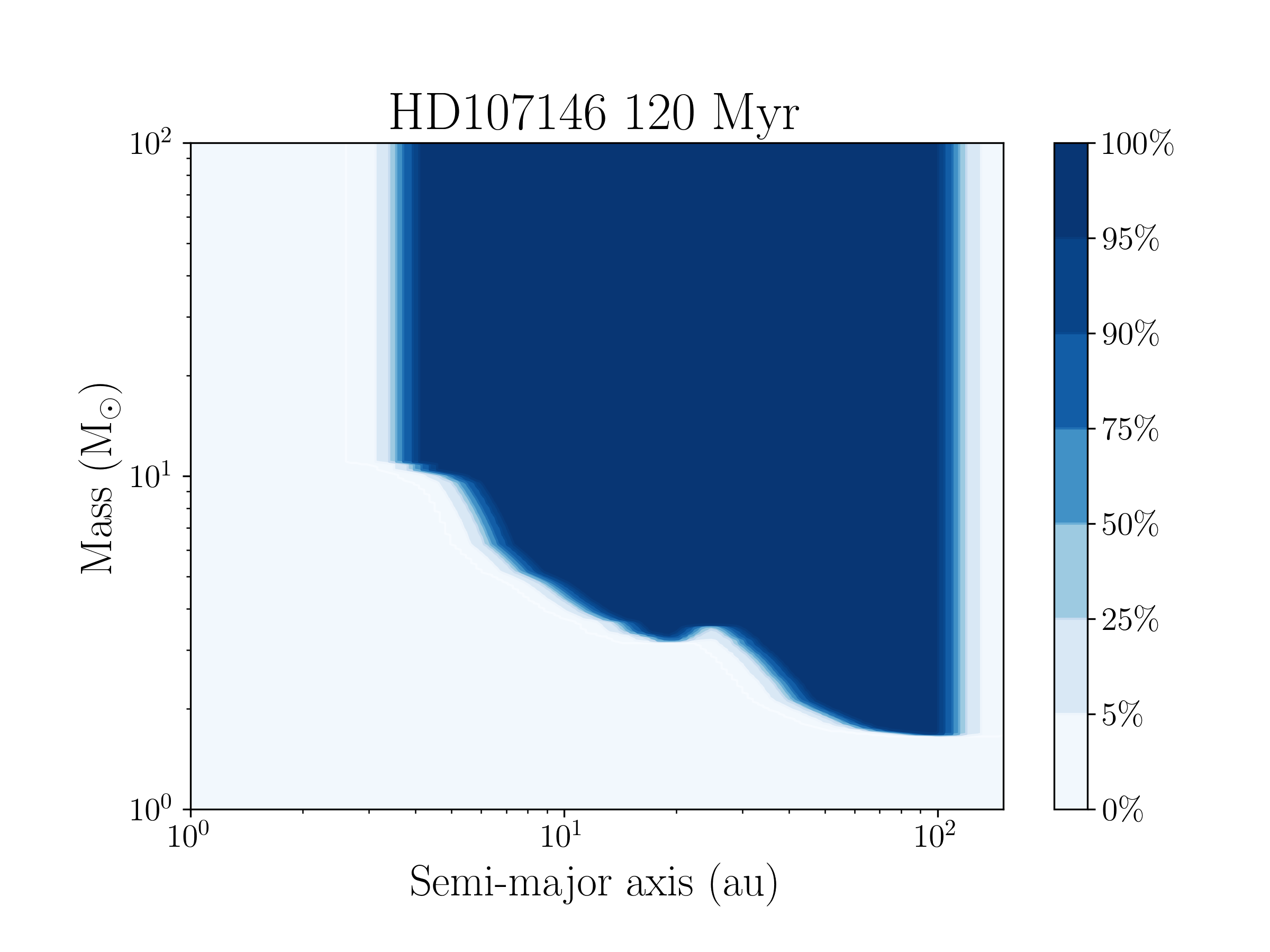}    
    \includegraphics[height=5.75cm]{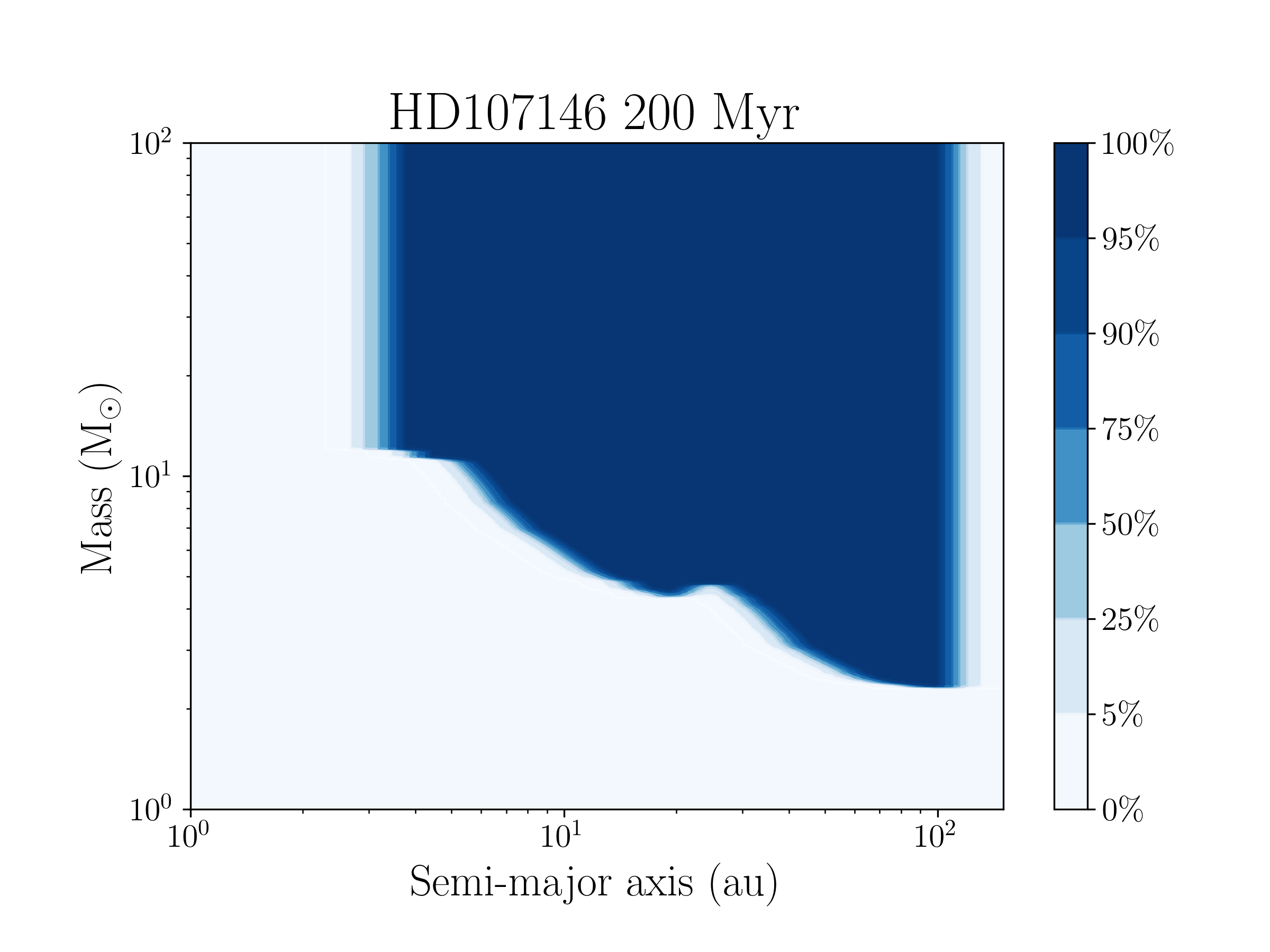} 
    \caption{Detection probability maps, obtained with the DMC code, for
      HD\,92945 (left panels) and HD\,107146 (right panels). Given the
      uncertainty on the age of the targets, we show the results of the runs
      performed using the contrast limits converted to mass using both the
      minimum and maximum age estimates (upper and lower panels, respectively).}
    \label{fig:DImaps}
\end{figure*}

The Exoplanet Detection Map Calculator \citep[Exo-DMC][]{Bonavita:2020ascl} is the latest (and for the first time in Python) rendition of the MESS \citep[Multi-purpose Exoplanet Simulation System][]{bonavita2012}, a Monte Carlo tool for the statistical analysis of direct imaging  survey results. 
In a similar fashion to its predecessors, the DMC combines the information on the target stars with the instrument detection limits to estimate the probability of detection of a given synthetic planet population, ultimately generating detection probability maps. \par 
For each star in the sample the DMC produces a grid of masses and physical separations of synthetic companions, then estimates the probability of detection given the provided detection limits. The default setup uses a flat distributions in log space for both the mass and semi-major axis but, in a similar fashion to its predecessors, the DMC allows for a high level of flexibility in terms of possible assumptions on the synthetic planet population to be used for the determination of the detection probability.  

For each point in the mass/semi-major axis grid the DMC generates a fixed number of sets of orbital parameters. By default all the orbital parameters are uniformly distributed except for the eccentricity, which is generated using a Gaussian eccentricity distribution with $\mu =0$ and $\sigma = 0.3$, following the approach by \cite{hogg2010} \citep[see][for details]{bonavita2013}. 
For this reason we were able to properly take into account the effects of projection when estimating the detection probability using the contrast limits in Fig.~\ref{f:HD92945contrast} and Fig.~\ref{f:HD107146contrast}. The DMC in fact calculates the projected separations corresponding to each orbital set for all the values of the semi-major axis in the grid \citep[see][for a detailed description of the method used for the projection]{bonavita2012}. This allows to estimate the probability of each synthetic companion to truly be in the instrument FoV and therefore to be detected, if the value of the mass is higher than the limit.

In this specific case we chose to restrict the inclination and the longitude of the node of each orbital set to make sure that all companions in the population would lie in the same orbital plane as the disk. This is equivalent to assuming there would be no strong misalignments ($\leq 5\deg$) between the disks and possible embedded planets. A massive misaligned planet would in fact reorient the disk through secular interactions and the new mid plane would be aligned with the planet. This would occur typically in timescales shorter than the age of these systems \citep{PearceWyatt2014}.\\
Moreover, since in both cases the disk does not appear to be particularly
eccentric \citep{2018MNRAS.479.5423M,2019MNRAS.484.1257M}, which would be the
case if the orbit of the embedded planets were \citep{PearceWyatt2014,
  2015MNRAS.453.3329P}, we assumed a sigma of 0.1 for the eccentricity
distribution. \\
In Figure~\ref{fig:DImaps} we show the results for both targets. In order to
take into account the uncertainty on the age values, two different DMC runs were
performed for each target, using the minimum and maximum values of the mass
limits, respectively. 

\subsubsection{Comparison with available radial velocity measurements}
\label{s:rvdmc}
%\begin{figure}
%    \centering
%    \includegraphics[height=5.75cm]{rvlim.png}
%    \caption{95\% detection limits obtained with the RV-DMC for HD~92945 (blue line) and HD\,107146 (orange line)}
%    \label{fig:RVlimits}
%\end{figure}

\begin{figure*}
    \centering
    \includegraphics[height=5.75cm]{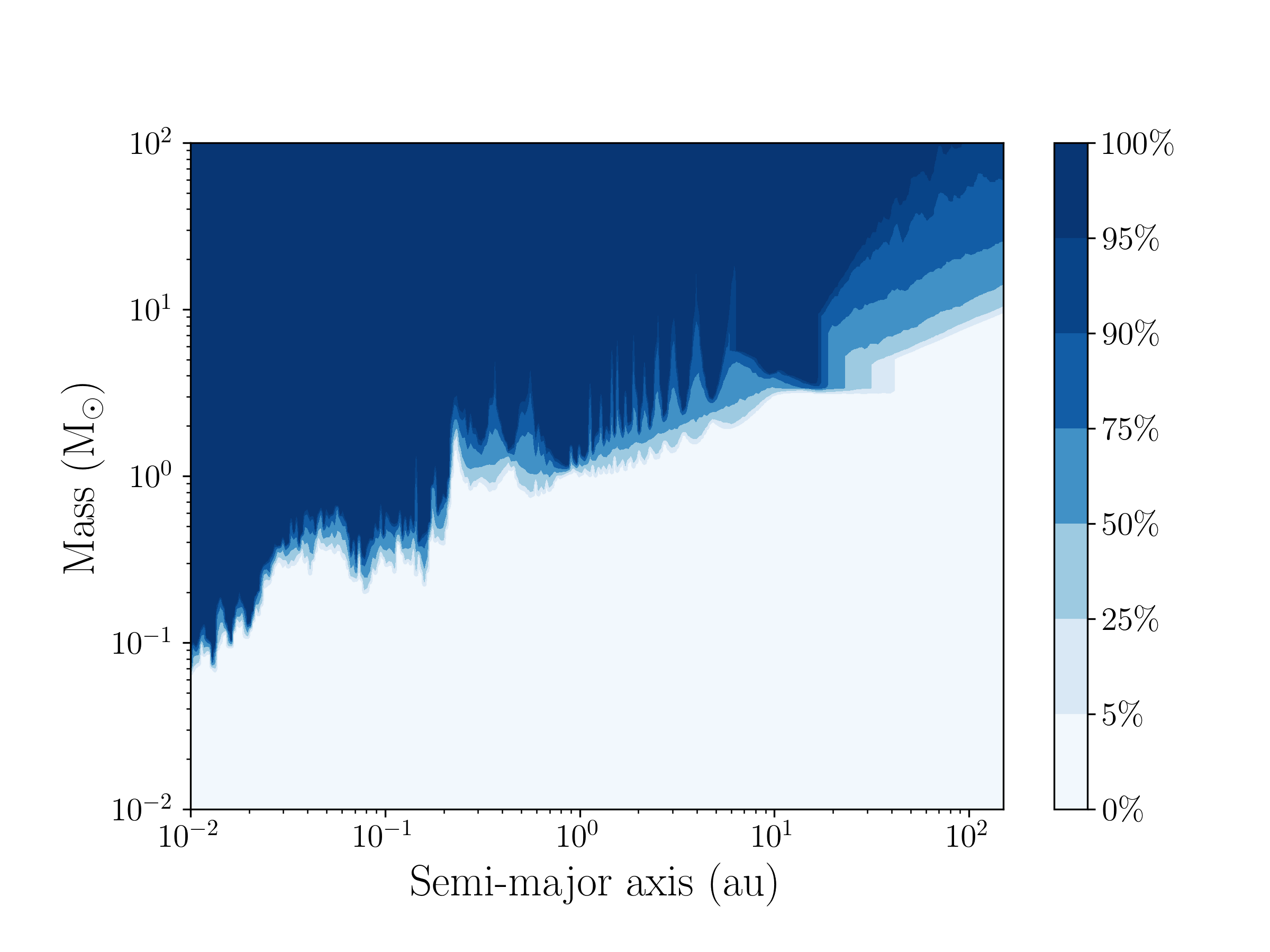}
    \includegraphics[height=5.75cm]{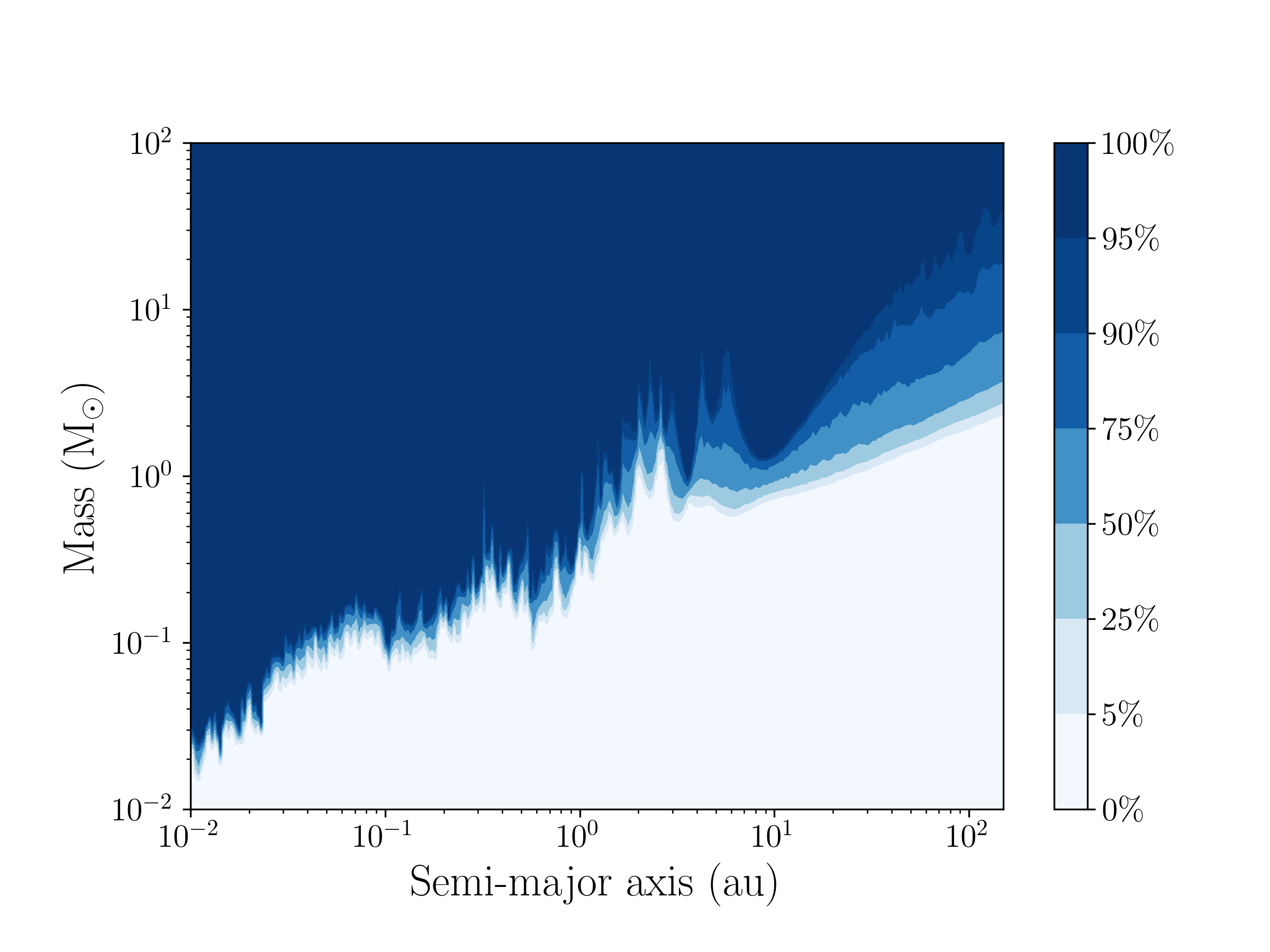}
    \caption{Combined RV + DI detection probability maps for for HD\,92945 (left
      panels) and HD\,107146 (right panels), estimated considering the best
      value of the probability for each point in the grid
      (see Sec.~\ref{s:rvdmc} for details). The values of the
      age for the magnitude to mass conversion of the DI limits in this case
      were set to 170~Myr for HD\,92945 and 150~Myr for HD\,107146.}
    \label{fig:comb_probmap}
\end{figure*}

While MESS was limited in its use to Direct Imaging data\footnote{With the exception of its MESS2 incarnation, which was designed to produce combined RV+DI detection probability maps \citep[see][for details]{lannier2017}}, the DMC can also be used to draw similar constraints using other kind of data sets, including radial velocity (RV) ones. 
Given the provided RV time series, the DMC uses the Local Power Analysis (LPA) approach described by \cite{meunier2012} to estimate, for each mass and separation in the grid, for what fraction of the generated orbital sets the signal generated by the companion would be compatible with the data.

Once the computation is completed, the code outputs a 2D detection probability map analogous to the ones produced with the DI module as well as a 1D detection limit showing the minimum value of the companion mass detectable with a given confidence level. The latter, which is roughly equivalent to extracting a specific detection probability contour from the 2D map, is the standard approach used to obtain RV detection limits in the past. The advantage of retaining the 2D information is that the resulting RV probability map can then be easily combined with the output of the DI module\footnote{One should note that the current version of DMC does not allow for a truly accurate estimate of the sensitivity achieved combining the two data sets, like the one performed in \cite{lannier2017}. The method used only allows for a rough estimate of the combined probability. A new module analogous to the MESS2 will be available in the future.}. \par 
We therefore run the RV-DMC on the available data for our two targets, using the same mass/semi-major axis grid as well as the same assumptions for the orbital parameter generation. 
The two sets of maps were then combined by considering, for each point in the grid, the best value of the probability. The resulting combined map, shown in Figure~\ref{fig:comb_probmap}, then contains the probability that a companion at a given mass and semi-major axis is detected using at least one of the two methods.

\subsubsection{Comparison with PMa results}

\begin{figure}
\centering
\includegraphics[width=\columnwidth]{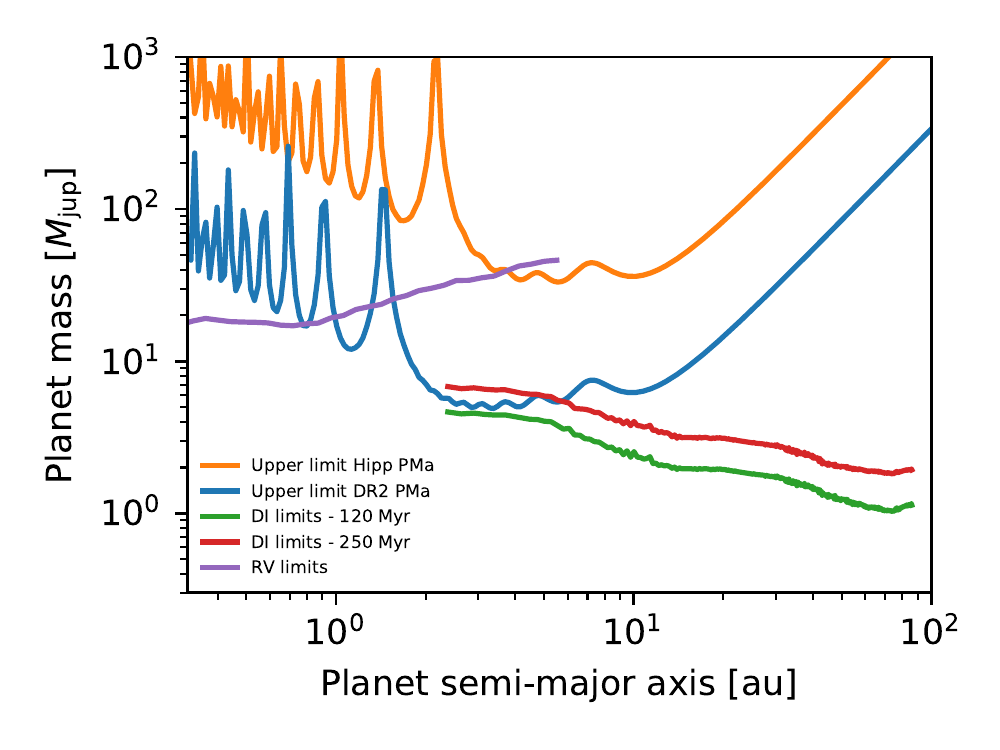}
\caption{Upper limits in mass obtained from the PMa measurements at
  Hipparcos epoch (orange line) and for the Gaia DR2 epoch (blue line)
  for HD\,92945 compared with the limits from
  the RV data (violet line, assuming a 95\% confidence level, see
  Sec.~\ref{s:rvdmc} for details) and from the DI data for an age of 120~Myr
  (green line) and for an age of 250~Myr (red line).}
\label{f:HD92945pma}
\end{figure}

\begin{figure}
\centering
\includegraphics[width=\columnwidth]{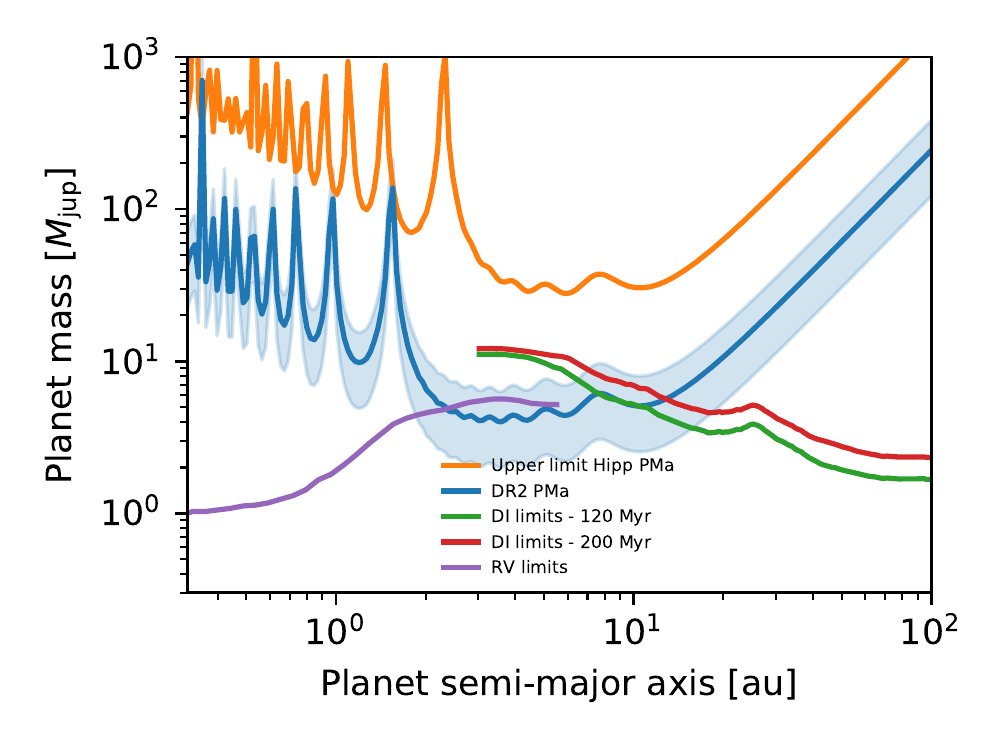}
\caption{Mass as function of the separation from the host star of the
  companion needed to explain the PMa measurement at the Gaia DR2 epoch
  for HD\,107146 (blue line).
  The blue shaded area display the 99.7\% confidence interval.
  The orange line represents the upper limits from PM measurements at
    the Hipparcos epoch while
  the violet line displays the mass limits from RV data (assuming a 95\% confidence level, see Sec.~\ref{s:rvdmc} for details). The DI mass limits
  adopting an age of 120~Myr are shown with the green line while the DI limits
  adopting an age of 200~Myr are shown by the red line.}
\label{f:HD107146pma}
\end{figure}

As explained in Section~\ref{s:datadeltamu} PMa data are available for both
targets and they can be used to calculate the mass of a possible object
  that cause the anomaly as function of the separation from the host star
  exploiting Equation~15 in \citep{2019A&A...623A..72K}. For HD\,92945 no
clear evidence of the presence of a companion emerges from these data. We can
then use these data just to put upper limits on the masses of possible
  companions. They are however comparable with those obtained through RV
  and DI only at separations between 1 and 10~au while they are clearly worse at
  lower and larger separations. This can be easily seen in
Figure~\ref{f:HD92945pma} where the upper limits from the PMa
measurements obtained using data from the Hipparcos and the Gaia DR2
epochs (orange and blue lines; labeled as upper limit Hipp PMa and upper
limit DR2 PMa) are compared with mass limits from RV data
(violet line) and from DI data (red and green lines). \par
Different is the situation for HD\,107146 where a 3.4$\sigma$ PMa obtained
  at the Gaia DR2 epoch strongly suggests the presence of a companion. Using
Equation~15 in \citet{2019A&A...623A..72K} and a Monte-Carlo approach as in
\citet{2020MNRAS.498.1319M} we can estimate the mass of the
companion responsible for the PMa, as a function of the semi-major axis.
This is represented by the blue line in Figure~\ref{f:HD107146pma} and is
  labeled as DR2 PMa while the shaded blue area surrounding this line is the
99.7\% confidence interval. The 95\% 1D  mass limits obtained with the RV module of Exo-DMC (violet line) combined with the mass limits from the DI data (red and green lines) allow to exclude large part of the possible
masses and separations for this putative object. In this way, if it exists, it
can just be at separations between $\sim$2-7~au and with a mass of 2-5~\MJup.
Future imaging instruments operating at extremely large telescopes (ELT) should
be able to image such a planet that will also be clearly detected through the
next Gaia data releases. It has to be noted that the presence of an asteroid
belt at a separation between 5 and 15~au has been inferred thanks to Spitzer
data \citep{2011ApJ...730L..29M,2014MNRAS.444.3164K}. If confirmed, the proposed
companion would lay just inside the position of this belt and its presence
could explain the formation of the belt with the pressure bump due to the
presence of such massive planet.
  
\section{Conclusion}
\label{s:conclusion}

We presented the results of the SPHERE observations of the stars HD\,92945
and HD\,107146. Both these stars are known to host a debris disk with an
outer planetesimals belt containing a wide gap. Moreover, they share similar
ages. Given the morphology of these gaps, it was proposed that these could have been carved by one or more planetary mass companion, residing
inside the gap for both systems \citep{2015ApJ...798..124R, 2018MNRAS.479.5423M} or possibly interior to the disk for HD~92945 since its gap appears marginally asymmetric \citep{2019MNRAS.484.1257M}. The
capability to put strong constraints on the mass of possible companions around
these objects can help to better understand the formation mechanisms of these
structures. Moreover, due to the paucity of gas in these disks
\citep{2018MNRAS.479.5423M,2019MNRAS.484.1257M}, the problem
of strong absorption that can be experienced for younger and gas-rich
disks \citep[like e.g. the case of HD\,163296; ][]{2019MNRAS.488...37M} should
be less important making more reliable the limits obtained around these objects.
On the other hand, the older ages of these objects with respect to systems
hosting protoplanetary disks makes harder to detect low mass planets. \par
In any case, the mass limits that we obtain from our observations are
much lower than obtained before with previous DI observations. Indeed, for both
the systems considered here, we were able to put very tight mass limits of
1-2~\MJup according to the age considered for the system. For what concerns
the inner region with respect to the planetesimal belt, we can reach for both
systems mass limits lower than 4-5~\MJup for the range of possible age
considered. These limits are enough to exclude some of the proposed
configurations of inner planets that could explain the structure of the
gap. \par
While the contrast obtained with SPHERE is still not enough to confirm or to
exclude large part of the range of possible companion masses that could explain
the shape of the planetesimal belt around HD\,92945 and HD\,107146, the mass
limits much tighter than those obtained in past survey, can help in limiting
the possible configurations of these planetary systems. This data demonstrate
the potentiality of the direct imaging in the understanding of the structure
of planetary systems. To fully exploit this potentiality, however, we will
probably need to wait for future instrumentation reaching higher contrasts and
larger spatial resolution. As an example the external planets responsible of
carving the gap should be recoverable using the JWST NirCAM operating in
coronagraphic observing mode in the wavelength range 3-5~\mic as at separations
of some tens of au it should be able to image companions down to the Saturn
mass \citep{2012SPIE.8442E..2NB}. \par
At low separations from the star, DI observations are not able to put tight
mass limits. Moreover, the presence of the coronagraph hides the inner region
of the planetary system. We then used both PMa and RV data to obtain
informations on the inner regions of these systems. As for HD\,92945, no
clear signal is obtained from PMa and we can just put upper limits that are
however comparable to those obtained through RV data (some tens of \MJup)
and through DI at higher separations. Much more interesting is the situation
for HD\,107146 where there is a tentative 3.4$\sigma$ PMa indicative of the
presence of a companion. Coupling this result with the limits coming from RV
and DI we are able to constrain both the separation ($\sim$2-6~au) and the
mass (2-5~\MJup) of such object. We note here that future ELT coronagraphic
instruments will possess the angular resolution requested to detect such an
object while the next Gaia releases should be able to further confirm its
existence.

\section*{Acknowledgments}
%The authors thanks the anonymous referee for the constructive comments that
%helped to strongly improve the quality of the present work. \par
Based on observations collected at the European Organisation for Astronomical Research in the Southern Hemisphere under ESO programme 095.C-0374(A), 1100.C-0481(D), 074.C-0037(A), 075.C-0202(A) and 192.C-0224(B).% and 0101.C-0843(A).
Partly Based on observations collected with the SOPHIE spectrograph on the 1.93 m telescope at Observatoire de Haute-Provence (CNRS), France under program 07A.PNP.CONS, 08A.PNP.CONS, 12A.DISC.WATS, 13A.PNP.DELF, 14A.PNP.LAGR, 14B.PNP.LAGR.
SPHERE is an instrument designed and built by a consortium
consisting of IPAG (Grenoble, France), MPIA (Heidelberg, Germany), LAM
(Marseille, France), LESIA (Paris, France), Laboratoire Lagrange
(Nice, France), INAF–Osservatorio di Padova (Italy), Observatoire de
Genève (Switzerland), ETH Zurich (Switzerland), NOVA (Netherlands),
ONERA (France) and ASTRON (Netherlands) in collaboration with
ESO. SPHERE was funded by ESO, with additional contributions from CNRS
(France), MPIA (Germany), INAF (Italy), FINES (Switzerland) and NOVA
(Netherlands).  SPHERE also received funding from the European
Commission Sixth and Seventh Framework Programmes as part of the
Optical Infrared Coordination Network for Astronomy (OPTICON) under
grant number RII3-Ct-2004-001566 for FP6 (2004–2008), grant number
226604 for FP7 (2009–2012) and grant number 312430 for FP7
(2013–2016). We also acknowledge financial support from the Programme National
de Plan\'{e}tologie (PNP) and the Programme National de Physique Stellaire
(PNPS) of CNRS-INSU in France. This work has also been supported by a grant from
the French Labex OSUG@2020 (Investissements d’avenir – ANR10 LABX56).
The project is supported by CNRS, by the Agence Nationale de la
Recherche (ANR-14-CE33-0018). It has also been carried out within the frame of
the National Centre for Competence in  Research PlanetS supported by the Swiss
National Science Foundation (SNSF). \par
This work has made use of data from the European Space Agency (ESA)
  mission {\it Gaia} (\url{https://www.cosmos.esa.int/gaia}), processed by
  the {\it Gaia} Data Processing and Analysis Consortium (DPAC,
  \url{https://www.cosmos.esa.int/web/gaia/dpac/consortium}). Funding for
  the DPAC has been provided by national institutions, in particular the
  institutions participating in the {\it Gaia} Multilateral Agreement.
This research has made use of the SIMBAD database, operated at CDS,
Strasbourg, France. \par
D.M., V.D.O., R.G., S.D. acknowledge support from the ``Progetti Premiali'' funding scheme of the Italian Ministry of Education, University, and Research and from the ASI-INAF agreement n.2018-16-HH.0. S.M. is supported by a research fellowship from Jesus College, University of Cambridge. M.B. acknowledges funding by the UK Science and Technology Facilities Council (STFC) grant no. ST/M001229/1.
T.H. acknowledges support from the
European Research Council under the Horizon 2020 Framework Program via the ERC
Advanced Grant Origins 83 24 28.

\section*{Data Availability}
The data underlying this article will be shared on reasonable request to the
corresponding author.
 
\bibliographystyle{mnras}
\bibliography{debris}

{\it
  $^{1}$INAF-Osservatorio Astronomico di Padova, Vicolo dell'Osservatorio 5, Padova, Italy, 35122-I\\
  $^{2}$Max Planck Institute for Astronomy, K\"onigstuhl 17, 69117 Heidelberg, Germany\\
  $^{3}$Institute of Astronomy, University of Cambridge, Madingley Road, Cambridge CB3 0HA, UK\\
  $^{4}$SUPA, Institute for Astronomy, University of Edinburgh, Blackford Hill, Edinburgh EH9 3HJ, UK\\
  $^{5}$Centre for Exoplanet Science, University of Edinburgh, Edinburgh EH9 3HJ, UK\\
  $^{6}$Dipartimento di Fisica a Astronomia "G. Galilei", Universit'a di Padova, Via Marzolo, 8, 35121 Padova, Italy \\
  $^{7}$Center for Space and Habitability, University of Bern, Gesellschaftsstrasse 6, 3012 Bern, Switzerland\\
  $^{8}$Universidad de Santiago de Chile, Av. Libertador Bernardo O’Higgins 3363, Estacion Central, Santiago, Chile\\
  $^{9}$School of Physics and Astronomy, Monash University, Clayton campus, VIC 3800, Australia \\
  $^{10}$ETH Zurich, Institute for Particle Physics and Astrophysics, Wolfgang-Pauli-Strasse 27, CH-8093 Zurich, Switzerland \\
  $^{11}$Department of Astronomy, Stockholm University, Stockholm, Sweden\\
  $^{12}$LESIA, Observatoire de Paris, Universit\'e PSL, CNRS, Sorbonne Universit\'e, Univ. Paris Diderot, Sorbonne Paris Cit\'e, \\ 5 place Jules Janssen, 92195 Meudon, France \\
  $^{13}$Univ. Lyon, Univ. Lyon 1, ENS de Lyon, CNRS, CRAL UMR 5574, 69230 Saint-Genis-Laval, France \\
  $^{14}$Aix Marseille Univ., CNRS, CNES, LAM, Marseille, France \\
  $^{15}$INAF-Catania Astrophysical Observatory, Via S. Sofia, 78, 95123 Catania, Italy \\
  $^{16}$Univ. Grenoble Alpes, CNRS, IPAG, 38000 Grenoble, France\\
  $^{17}$Konkoly Observatory, Research Centre for Astronomy and Earth Sciences, Konkoly-Thege Mikl\'os \'ut 15-17, H-1121 Budapest, Hungary\\
  $^{18}$Instituto de Física y Astronomía, Facultad de Ciencias, Universidad de Valparaíso, Av. Gran Bretaña 1111, Playa Ancha, Valparaíso, Chile \\
  $^{19}$N\'ucleo Milenio Formaci\'on Planetaria - NPF, Universidad de Valpara\'iso, Av. Gran Bretana 1111, Playa Ancha, Valpara\'iso, Chile \\
  $^{20}$Geneva Observatory, University of Geneva, Chemin des Maillettes 51, CH-1290 Sauverny, Switzerland \\
  $^{21}$European Space Agency (ESA), ESA Office, Space Telescope Science Institute, 3700 San Martin Drive, Baltimore, MD 21218, USA \\
  $^{22}$European Southern Observatory, Alonso de Còrdova 3107, Vitacura, Casilla 19001, Santiago, Chile \\
  $^{23}$Nucleo de Astronomia, Facultad de Ingenieria y Ciencias, Universidad Diego Portales, Av. Ejercito 441, Santiago, Chile \\
  $^{24}$Escuela de Ingenieria Industrial, Facultad de Ingenieria y Ciencias, Universidad Diego Portales, Av. Ejercito 441, Santiago, Chile \\
  $^{25}$Departamento de Astronom\'ia, Universidad de Chile, Casilla 36-D, Santiago \\
  $^{26}$INAF - Osservatorio Astronomico di Capodimonte, Via Salita Moiariello 16, 80131 Napoli, Italy \\
  $^{27}$STAR Institute, Universit\'e de Liege, Allee du Six Août 19c, 4000 Liege, Belgium \\
  $^{28}$University of Michigan, Astronomy Department, USA }

%\clearpage
\appendix

\label{lastpage}

\end{document}